\documentclass{jpp}

\usepackage{graphicx}

\usepackage{natbib}
\usepackage{amsmath}
\usepackage{amssymb}
\usepackage{color}
\usepackage{bm}
\usepackage{authblk}

\ifCUPmtlplainloaded \else
  \checkfont{eurm10}
  \iffontfound
    \IfFileExists{upmath.sty}
      {\typeout{^^JFound AMS Euler Roman fonts on the system,
                   using the 'upmath' package.^^J}%
       \usepackage{upmath}}
      {\typeout{^^JFound AMS Euler Roman fonts on the system, but you
                   dont seem to have the}%
       \typeout{'upmath' package installed. JPP.cls can take advantage
                 of these fonts, if you use 'upmath' package.^^J}%
      }
  \else
  \fi
\fi

\ifCUPmtlplainloaded \else
  \checkfont{msam10}
  \iffontfound
    \IfFileExists{amssymb.sty}
      {\typeout{^^JFound AMS Symbol fonts on the system, using the
                'amssymb' package.^^J}%
       \usepackage{amssymb}%
       \let\le=\leqslant  
       \let\ge=\geqslant  
      }{}
  \fi
\fi

\ifCUPmtlplainloaded \else
  \checkfont{msam10}
  \iffontfound
    \IfFileExists{amssymb.sty}
      {\typeout{^^JFound AMS Symbol fonts on the system, using the
                'amssymb' package.^^J}%
       \usepackage{amssymb}%
       \let\le=\leqslant  
       \let\ge=\geqslant  
      }{}
  \fi
\fi

\ifCUPmtlplainloaded \else
\IfFileExists{amsbsy.sty}
             {\typeout{^^JFound the 'amsbsy' package on the system, using it.^^J}%
     \usepackage{amsbsy}}
    {}
\fi

\newsavebox{\astrutbox}
\sbox{\astrutbox}{\rule[-5pt]{0pt}{20pt}}

\newcommand{\Alfven}{Alfv\'{e}n }
\newcommand{\Alfvenic}{Alfv\'{e}nic }
\newcommand{\T}[1]{{\tt #1}}
\newcommand{\V}[1]{\mathbf{#1}}

\newcommand{\zhat}{\mbox{$\hat{\mathbf{z}}$}}

\newcommand{\figref}[1]{Fig.~\ref{#1}}   
\newcommand{\secref}[1]{\S\ref{#1}}                                                                                        

\title[Energy transfer in \Alfven wavepacket collisions]{The \Alfvenic nature of energy transfer mediation in localized, strongly nonlinear \Alfven wavepacket collisions}

\author[1]{J.~L.~Verniero}
\author[2]{G.~G.~Howes}

\affil[1]{Department of Mathematics, University of Iowa, Iowa City IA 54224, USA}
\affil[2]{Department of Physics and Astronomy,University of Iowa, Iowa City IA 54224, USA}
\pubyear{2010}
\volume{650}
\pagerange{119--126}
\date{?; revised ?; accepted ?. - To be entered by editorial office}

\begin{document}
\maketitle

\begin{abstract}
In space and astrophysical plasmas, violent events or instabilities inject energy into turbulent motions at large scales.  Nonlinear interactions among the turbulent fluctuations drive a cascade of energy to small perpendicular scales at which the energy is ultimately converted into plasma heat. Previous work with the incompressible magnetohydrodynamic (MHD) equations has shown that this turbulent energy cascade is driven by the nonlinear interaction between counterpropagating \Alfven waves -- also known as \Alfven wave collisions. Direct numerical simulations of weakly collisional plasma turbulence enables deeper insight into the nature of the nonlinear interactions underlying the turbulent cascade of energy. In this paper, we directly compare four cases: both periodic and localized \Alfven wave collisions in the weakly and strongly nonlinear limits. Our results reveal that in the more realistic case of localized \Alfven wave collisions (rather than the periodic case), all nonlinearly generated fluctuations are \Alfven waves, which mediates nonlinear energy transfer to smaller perpendicular scales.
\end{abstract}



\section{Introduction}
\label{sec:intro}
 
 	Turbulence plays a vital role in the dynamics of space plasmas such as the solar wind, astrophysical plasma systems such as galaxy clusters, and laboratory plasma environments such as magnetically confined fusion plasmas. Driven by violent events or instabilities at a large scale (such as impulsive magnetic reconnection in active regions on the Sun), turbulent energy is transferred to smaller perpendicular scales and eventually turned into plasma heat via dissipative mechanisms.  Understanding the entire cascade of turbulent energy and how it converts turbulent energy into plasma heat is crucial for understanding how poorly understood astrophysical, space, and laboratory plasma systems evolve. For this reason, the dynamics of the turbulent energy transfer remains a fervent research topic of plasma physics.
 
 	In contrast to the eddies that describe hydrodynamic turbulence, \Alfven waves -- waves supported by magnetic tension that propagate up or down along the magnetic field -- dominate the physics of turbulent motions in a magnetized plasma, a concept first proposed by early research on incompressible magnetohydrodynamic (MHD) turbulence in the 1960s \citep{Iroshnikov:1963,Kraichnan:1965}. Formulating the picture of plasma turbulence in this way, \Alfven wave collisions are known as the ``fundamental building block of plasma turbulence" \citep{Kraichnan:1965,Howes:2013a}. Hence, studying the details of the nonlinear energy transfer of \Alfven wave collisions lays important groundwork for understanding the turbulent energy cascade within a fully turbulent medium where these \Alfven wave collisions are omnipresent.  Following significant work on incompressible MHD turbulence \citep{Sridhar:1994,Montgomery:1995,Ng:1996,Galtier:2000},  a recent study has computed an analytical solution for the evolution of \Alfven wave collisions in the weakly nonlinear limit \citep{Howes:2013a} which has been validated by nonlinear gyrokinetic numerical simulations \citep{Nielson:2013a} and verified in the laboratory  \citep{Howes:2012b,Howes:2013b,Drake:2013}.  
	
	As described in detail in \citet{Howes:2013a}, the general picture of nonlinear energy transfer in the weakly nonlinear case is as follows. \Alfven wave modes are of the form $\hat{\mathbf{k}}=(k_x/k_{\perp 0}, k_y/k_{\perp 0}, k_z/k_{\parallel0})$, where $k_{\perp 0}$ and $k_{\parallel 0}$ are the perpendicular and parallel wave numbers relative to the equilibrium magnetic field direction of the initial two \Alfven waves in the MHD limit, $k_\perp \rho_i \ll 1$. First, the perpendicularly polarized primary \Alfven wave modes $\hat{\mathbf{k}}_1^-$ = (1,0,1) and 
$\hat{\mathbf{k}}_1^+$ = (0,1,-1) interact nonlinearly to give $\hat{\mathbf{k}}_1^-$+$\hat{\mathbf{k}}_1^+$ = $\hat{\mathbf{k}}_2^{(0)}$ = (1,1,0). Under the periodic conditions adopted to facilitate an analytical solution, the secondary mode is a purely magnetic fluctuation, physically representing a shear in the magnetic field which oscillates at a rate of $2\omega_A$, where $\omega_A \equiv k_{\parallel 0} v_A$ is the frequency of the two primary \Alfven waves. This inherently nonlinear mode has no parallel variation ($k_\parallel = 0$), therefore it is not an \Alfven mode since it does not satisfy the \Alfven wave dispersion relation, $\omega= k_\parallel v_A$. In other words, this $\hat{\mathbf{k}}_2^{(0)}$ mode does not propagate as an \Alfven wave, which would have a parallel phase velocity $\omega/k_\parallel= v_A$ and a parallel group velocity $\partial \omega/\partial k_\parallel= v_A$.  Furthermore, the amplitude of this secondary mode rises and falls in an oscillatory fashion at a frequency of $2 \omega_A$, never gaining energy secularly.  This secondary mode is essentially a nonlinearly generated beat mode \citep{Drake:2016}.  Next, each primary mode $\hat{\mathbf{k}}_1^\pm$ interacts with this secondary mode 
$\hat{\mathbf{k}}_2^{(0)}$ to transfer energy secularly to two tertiary modes, $\hat{\mathbf{k}}_1^\pm+ \hat{\mathbf{k}}_1^{(0)} = \hat{\mathbf{k}}_3^\pm$, where $\hat{\mathbf{k}}_3^- = (2,1,1)$ and 
$\hat{\mathbf{k}}_3^+ = (1,2,-1)$. These tertiary modes $\hat{\mathbf{k}}_3^\pm$ have the same value of $k_\parallel$ as the corresponding primary modes $\hat{\mathbf{k}}_1^\pm$. The amplitude of these tertiary modes  $\hat{\mathbf{k}}_3^\pm$ grows secularly in time, with energy transfer from the primary modes $\hat{\mathbf{k}}_1^\pm$ mediated by the strictly oscillatory secondary mode $\hat{\mathbf{k}}_0^{(0)}$. The analytical calculation \citep{Howes:2013a} therefore identifies the key role of the nonlinearly generated secondary mode with $k_\parallel =0$ in the nonlinear transfer of energy from larger to smaller perpendicular scales relative to the background magnetic field. The purpose of the present study is to illuminate the nature of this secondary mode in the more realistic case of collisions between initially separated \Alfven wavepackets.
	
	Strongly nonlinear MHD plasma turbulence simulations have led to another important discovery about plasma turbulence, that intermittent current sheets develop \citep{Matthaeus:1980,Meneguzzi:1981} and turbulent energy dissipation is mostly concentrated within these sheets \citep{Uritsky:2010,Osman:2011,Zhdankin:2013}. Therefore, evidence for the connection between the development of current sheets and the dissipation of turbulent energy into plasma heat has been sought after observationally \citep{Osman:2011,Borovsky:2011,Osman:2012a,Perri:2012a,Wang:2013,Wu:2013,Osman:2014b} and numerically \citep{Wan:2012,Karimabadi:2013,TenBarge:2013a,Wu:2013,Zhdankin:2013}. Recent work has shown that, in the strong turbulence limit, \Alfven wave collisions generate current sheets \citep{Howes:2016b}, an important breakthrough connecting the self-consistent development of intermittent current sheets and the nonlinear mechanism responsible for transferring turbulent energy to smaller scales. Subsequent work using the new field-particle correlation technique 
\citep{Klein:2016a,Howes:2017a,Howes:2017c,Klein:2017b} has shown that the particle energization in these current sheets involves collisionless energy transfer via the Landau resonance \citep{Howes:2017d}.
	
	The previous work on \Alfven wave collisions \citep{Howes:2013a,Nielson:2013a} explored the nonlinear interactions between two perpendicularly polarized, counterpropagating plane \Alfven waves under periodic boundary conditions.  These two plane \Alfven waves were initially overlapping before they began to interact nonlinearly, an unrealistic, idealized set up that enabled an asymptotic analytical solution to be obtained in the weakly nonlinear limit. A depiction of the initial conditions in this case is shown in \figref{fig:setup}(a), where the variation along the direction $z$ (parallel to the equilibrium magnetic field) for each of the two initial, perpendicularly polarized \Alfven waves is plotted.  The upward propagating \Alfven wave has a $\delta B_y$ polarization with a perpendicular Fourier mode $(1,0)$ (blue) and the downward propagating \Alfven wave has a $\delta B_x$ polarization with a perpendicular Fourier mode $(0,1)$ (red).  Note these initial plane \Alfven wave modes fill the simulation domain and are periodic in both the perpendicular plane as well as the parallel direction.  We refer to this \Alfven wave initialization as the \emph{periodic case}.  Note that the periodic boundary conditions are not what makes this scenario unrealistic, but rather the fact that the two waves started on top of each other and consequently did not arrive in those positions while undergoing a self-consistent nonlinear interaction.
	
	An important question is whether the key properties of the nonlinear evolution of \Alfven wave collisions found in this idealized periodic case persists for the more realistic case of the interaction between two initially separated \Alfven wavepackets.  To answer this question, we perform nonlinear kinetic simulations of the interaction between two localized \Alfven wavepackets that do not initially overlap, as shown in \figref{fig:setup}(b). Here, the upward propagating \Alfven wave has a $\delta B_y$ polarization but the wavepacket is localized along the field parallel direction around $z = -L_z/4$.  Note that this \Alfven wavepacket remains periodic in the perpendicular plane, with its variation given by the Fourier mode $(1,0)$ (blue).  The downward propagating \Alfven wave has a $\delta B_x$ polarization, is localized in $z$ around $z = L_z/4$, and corresponds to a perpendicular Fourier mode $(0,1)$ (red).  Although the simulation domain itself is periodic in the $z$ direction, such that a wave propagating in the $+z$ direction will exit the domain  at $z=L_z/2$ and re-enter the domain at $z=-L_z/2$, the localization of the wavepackets along $z$ means that these two wavepackets will not interact nonlinearly until they come together and overlap along $z$, a more realistic situation.  We refer to this initially separated \Alfven wavepacket initialization as the \emph{localized case}.
	
	Our previous study of strongly nonlinear, localized \Alfven wave collisions \citep{Verniero:2017a} found that indeed nonlinear interactions between initially separated wavepackets facilitate the cascade of energy to smaller perpendicular scales relative to the background magnetic field and self-consistently give rise to current sheets, just as found in the periodic case.  But that study employed asymmetric initial \Alfven wavepackets (see Figure 1 of \citet{Verniero:2017a}), where one of the wavepackets had a significant $k_\parallel =0$ component initially relative to the background magnetic field.  Since it is the secondary mode with $k_\parallel=0$ that plays the key role in mediating the secular transfer of energy to smaller perpendicular scales in the periodic case, it is important to ensure that the non-zero $k_\parallel =0$ component of the wavepacket in \citet{Verniero:2017a} does not affect the results in a fundamental way.  To address this issue, we pursue here a detailed comparison of periodic \Alfven wave and localized \Alfven wavepacket collisions, where the initial wavepackets are symmetric and neither wavepacket has a significant $k_\parallel=0$ component.  This study will enable us to determine the nature of the nonlinearly generated modes that mediate the cascade of energy to smaller perpendicular scales relative to the background magnetic field in the localized case and to ensure that the non-zero $k_\parallel=0$ component in the \citet{Verniero:2017a} study did not qualitatively alter the resulting cascade by artificially initializing a mode that dominates the nonlinear energy transfer. 
	
	We aim to answer two primary questions: (i) What is the nature of the nonlinearly generated secondary mode that mediates the cascade of energy in localized \Alfven wave collisions?; and (ii) How does the localization of the interacting \Alfven waves into separated wavepackets affect the qualitative and quantitative evolution of the perpendicular cascade of energy and the development of current sheets?
	
	In \secref{sec:sim}, we describe the setup of the simulation for each of the four cases being compared. The nonlinear energy evolution of each case is presented in \secref{sec:energy}. Our results in \secref{sec:ver} show that the secondary (1,1) mode is an \Alfven wave mode.  The strongly and weakly nonlinear limits are compared in \secref{sec:sw}.  Current sheet development is confirmed in \secref{sec:cur}. Conclusions are discussed in \secref{sec:con}.

 \section{Simulation}
 \label{sec:sim}
  
  The nonlinear interaction between two counterpropagating localized \Alfven wavepackets or periodic 
  \Alfven waves is simulated using the Astrophysical Gryokinetics code \T{AstroGK} \citep{Numata:2010}.  \T{AstroGK} evolves the perturbed gyroaveraged distribution function
$h_s(x,y,z,\lambda,\varepsilon)$ for each species $s$, the scalar
potential $\varphi$, the parallel vector potential $A_\parallel$, and
the parallel magnetic field perturbation $\delta B_\parallel$
according to the gyrokinetic equation and the gyroaveraged Maxwell's
equations \citep{Frieman:1982,Howes:2006}. Velocity space coordinates
are $\lambda=v_\perp^2/v^2$ and $\varepsilon=v^2/2$. The domain is a
periodic box of size $L_{\perp }^2 \times L_{z}$, elongated along the
straight, uniform mean magnetic field $\mathbf{B}_0=B_0 \zhat$, where all
quantities may be rescaled to any parallel dimension satisfying $L_{z}
/L_{\perp } \gg 1$. Uniform Maxwellian equilibria for ions (protons)
and electrons are chosen, with a realistic mass ratio $m_i/m_e=1836$.
Spatial dimensions $(x,y)$ perpendicular to the mean field are treated
pseudospectrally; an upwind finite-difference scheme is used in the
parallel direction, $z$. Collisions employ a fully conservative,
linearized collision operator with energy diffusion and pitch-angle
scattering \citep{Abel:2008,Barnes:2009}.

  To reveal details of the turbulent transfer of energy through the interaction of \Alfven waves, we directly compare four simulations runs: 
 \begin{enumerate}
 \item Localized \Alfven wavepacket collisions in the strongly nonlinear limit, \T{LS}
 \item Periodic \Alfven wave collisions in the strongly nonlinear limit, \T{PS}
 \item Localized \Alfven wavepacket collisions in the weakly nonlinear limit, \T{LW}
 \item Periodic \Alfven wave collisions in the weakly nonlinear limit, \T{PW}
 \end{enumerate}

  \begin{figure}
  \hspace{0.1in} (a) Periodic case in strongly nonlinear limit at $t/T_c$=0  
\centering{ \includegraphics[trim=0 100bp 0 100bp,clip,scale=.5]{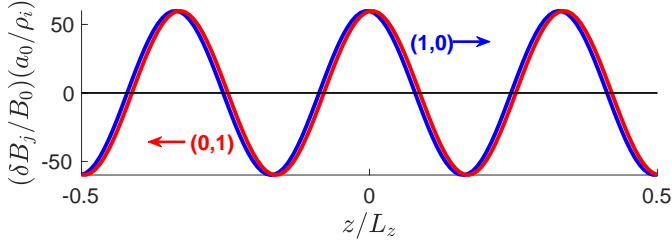}}
\vfill
\hspace{0.1in} (b) Localized case in strongly nonlinear limit at $t/T_c$=0
\centering{ \includegraphics[trim=0 100bp 0 100bp,clip,scale=.5]{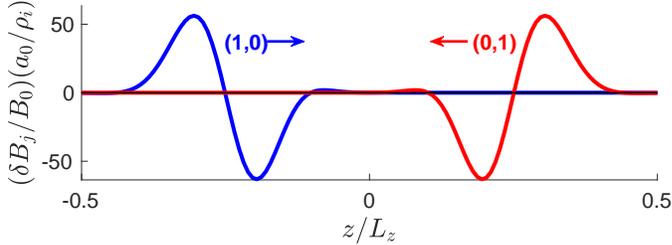}}
 \caption{Setup for perpendicularly polarized \Alfven waves in the localized and periodic cases. Note that the blue curve corresponds to the $(k_x,k_y)$ = (1,0) mode and the red curve corresponds to the $(k_x,k_y)$ = (0,1) mode.   Note that the blue and red fluctuations are polarized perpendicularly to each other, with $\delta B_x$ (red) and $\delta B_y$ (blue).
 \label{fig:setup}}
 \end{figure}

 For all cases, the plasma parameters are ion plasma beta $\beta_i$ = 1 and ion-to-electron temperature ratio $T_i/T_e$ = 1. We choose a perpendicular simulation domain size $L_{\perp}=40 \pi \rho_i$ with simulation resolution $(n_x,n_y,n_z,n_\lambda,n_\varepsilon,n_s)= (32,32,128,32,16,2)$ such that our initial \Alfven waves fall into the MHD limit, $k_\perp \rho_i \ll 1$. The fully resolved perpendicular range in this dealiased pseudospectral method covers $0.05 \le k_\perp \rho_i \le 0.5$. Here the ion thermal Larmor radius is $\rho_i= v_{ti}/\Omega_i$, the ion thermal velocity is $v_{ti}^2 =2T_i/m_i$, the ion cyclotron frequency is $\Omega_i= q_i B_0/(m_i c)$, and the temperature is given in energy units.  The parallel length of the simulation domain is $L_z$, extending over the range [$-L_z/2$,$L_z/2$].  Note that the simulation domain is triply periodic, so when a wavepacket exits the domain at $z=\pm L_z/2$, it re-enters at the opposite end at $z=\mp L_z/2$, enabling the two wavepackets to undergo successive collisions with each other.  The linearized Landau collision operator \citep{Abel:2008,Barnes:2009} is employed with collisional coefficients $\nu_i=\nu_e= 10^{-3}k_\parallel v_A$, yielding weakly collisional dynamics with $ \nu_s /\omega \ll 1$. 

The initial \Alfven wavepackets have perpendicular wave vectors 
$\mathbf{k}_\perp^- \rho_i = (k_x \rho_i,k_y \rho_i) = (0.05,0)$ for the upward ($z^-$) wavepacket and 
$\mathbf{k}_\perp^+ \rho_i = (k_x \rho_i,k_y \rho_i) = (0,0.05)$ for the downward ($z^+$) wavepacket, so both waves have the same initial perpendicular wavenumber $k_\perp^\pm \rho_i=0.05$, but are polarized perpendicular to each other. For brevity, we will refer to modes normalized to the domain scale perpendicular wave vector  
$k_{\perp 0} \equiv 2 \pi/ L_\perp$, giving  $\mathbf{k}_\perp^-/k_{\perp 0} =  (k_x /k_{\perp 0} ,k_y/ k_{\perp 0}) = (1,0)$.

\figref{fig:setup} illustrates the initial conditions for both the (a) periodic and (b) localized cases.  In panel (a), we plot the waveforms for the periodic cases, which are exactly the same as the localized case but without the application of the windowing function in $z$, so that the localized and periodic cases are directly comparable.  Here we plot the waveforms along the parallel direction $z$ at $t=0$ of the perpendicular Fourier mode $(k_x/k_{\perp 0}, k_y/k_{\perp 0}) = (1,0)$ of $\delta B_y$ (blue) and of the perpendicular Fourier mode $(0,1)$ of $\delta B_x$ (red) for the localized \Alfven wavepacket case in panel (b).   The localization along the $+z$ direction is specified using the procedure outlined in the Appendix~A of \citet{Verniero:2017a} with the parameters
$k_z a_0 = 3 , \delta = 0, z_0 = -\frac{\pi}{2} a_0 = -L_\parallel /4, \Delta_z = 1.2a_0$ and an exponent $p=2$.  For the wave which propagates in the $-z$ direction, the parameters are $k_z a_0 = -3 , \delta = 0, z_0 = \frac{\pi}{2} a_0 = L_\parallel /4, \Delta_z = 1.2a_0$ and an exponent $p=2$.  \figref{fig:setup} shows the amplitudes for the strongly nonlinear (a) periodic and (b) localized cases; the weakly nonlinear cases have the same initial waveforms but smaller amplitudes.

The amplitude of the initial wavepackets is parameterized by the nonlinearity parameter \citep{Goldreich:1995}, defined by taking the ratio of the magnitudes of the linear to the nonlinear terms in the incompressible MHD equations
 \citep{Howes:2013a,Nielson:2013a}. In terms of Elsasser variables, defined by
  $\V{z}^{\pm} = \V{u} \pm \delta \V{B}/\sqrt{4 \pi (n_{0i} m_i+n_{0e}m_e)}$, the nonlinearity parameter is defined by \newline
 $\chi^\pm \equiv |\V{z}^{\mp}\cdot \nabla \V{z}^{\pm}|/|\V{v}_A \cdot \nabla \V{z}^{\pm}|$, where $\chi^\pm$
characterizes the strength of the nonlinear distortion of the $\V{z}^{\pm}$ \Alfven wave by the counterpropagating $\V{z}^{\mp}$ \Alfven wave. For the particular initial \Alfven wavepackets shown in \figref{fig:setup}, the nonlinearity parameter simplifies to $\chi^\pm =2 k_\perp \delta B_\perp^\mp /(k_\parallel B_0)$. With the $\V{z}^{\pm}$ wavepackets having parallel wavenumbers of approximately $k_\parallel a_0= \mp 3$, where $a_0=L_z/2\pi$, the amplitude of the
wavepackets in the strongly nonlinear case $(\delta B_\perp^\pm /B_0) (a_0/\rho_i) \simeq 60$ gives $\chi^\pm=2$ and the amplitude of the wavepackets in the weakly nonlinear case $(\delta B_\perp^\pm /B_0) (a_0/\rho_i) \simeq 4$ gives $\chi^\pm=0.13$. Critically balanced, strong turbulence corresponds to a nonlinearity parameter of $\chi \sim 1$ \citep{Goldreich:1995}, and weak turbulence corresponds to $\chi \ll  1$, so these simulations fall into the desired limits of strong and weak nonlinearity, respectively.

 \section{Results}
 
The nonlinear evolution of the localized and periodic strong and weak \Alfven wave collisions during the first few collisions is concisely illustrated by a plot of the evolution of the energy in particular perpendicular Fourier modes in \figref{fig:energy3col}.  A meaningful quantitative comparison between the localized cases and the periodic cases is made possible by selecting comparable energies for each Fourier mode and a suitable definition of the \Alfven wave collision timescale in each case.

First, because the waveform in the $z$ direction differs between the localized and periodic cases, we choose to integrate the energy of each perpendicular Fourier mode $(k_x/k_{\perp 0},k_y/k_{\perp 0})$ along the $z$ direction to facilitate comparison.  

Second, we choose to normalize our timescales to the appropriate timescale of a single complete \Alfven wave collision in both the localized and periodic cases. In the localized case, the wavepackets collide twice during the time it takes for an \Alfven wave to propagate the parallel length of the domain, defined by $T_{L_z} \equiv L_z/v_A$.  By comparison, each wavelength in the periodic case passes through three wavelengths of the counterpropagating  waves during one wave-crossing period  $T_{L_z} $.  Therefore, we define the collision time as  $T_c^{(l)} = T_{L_z}/2$ for the localized \Alfven wavepacket collision case and $T_c^{(p)} = T_{L_z}/3$ for the periodic \Alfven wave collision case. To further illustrate the evolution in the localized case, note that the first collision begins when the counterpropagating wavepackets begin to overlap in $z$ at $t/T_c^{(l)}=1/6$ and ends at  $t/T_c^{(l)}=5/6$.  Subsequently, the second collision begins at $t/T_c^{(l)}=7/6$ and ends at  $t/T_c^{(l)}=11/6$.
 
 \subsection{Evolution of Energy of Secondary (1,1) mode}
 \label{sec:energy}
  
\begin{figure}
  \hspace{0.1in} (a) \T{LS} \hspace{2.7 in}  (b) \T{PS}
  \vfill
 \includegraphics[scale = .25]{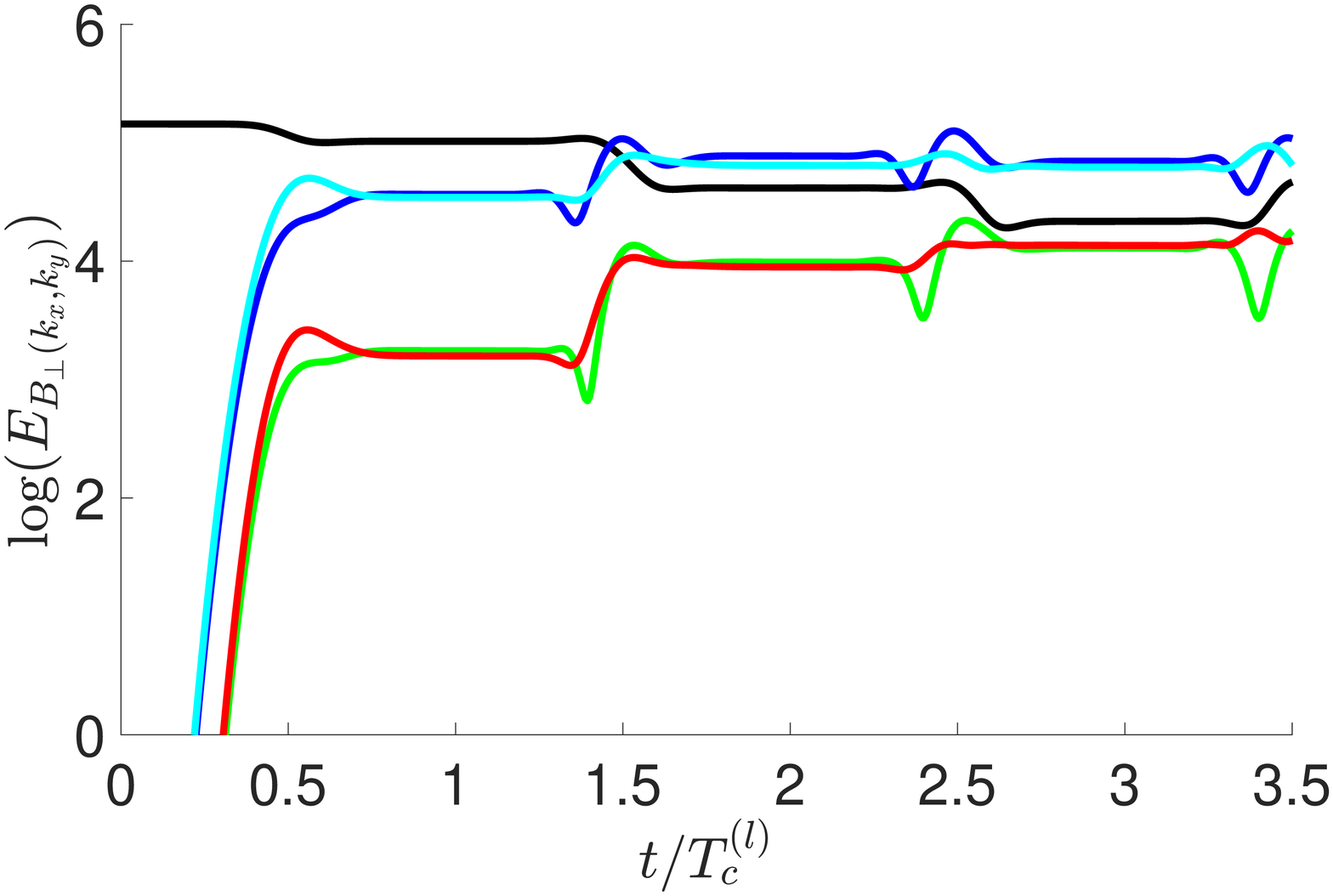}\hfill
 \includegraphics[scale = .25]{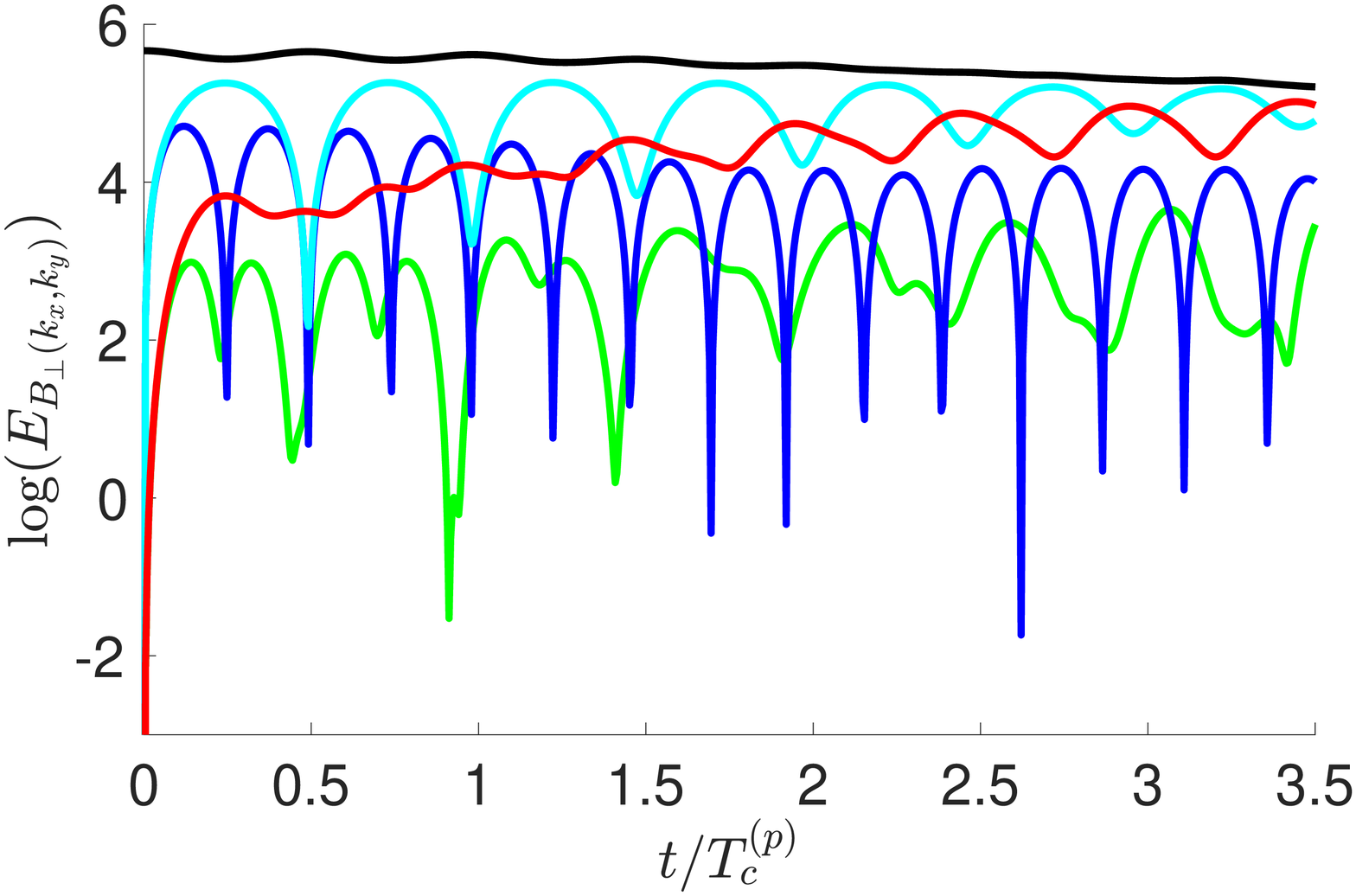}
 \vfill
 \hspace{0.1in} (c) \T{LW} \hspace{2.7 in}  (d) \T{PW}
 \vfill
 \includegraphics[scale = .25]{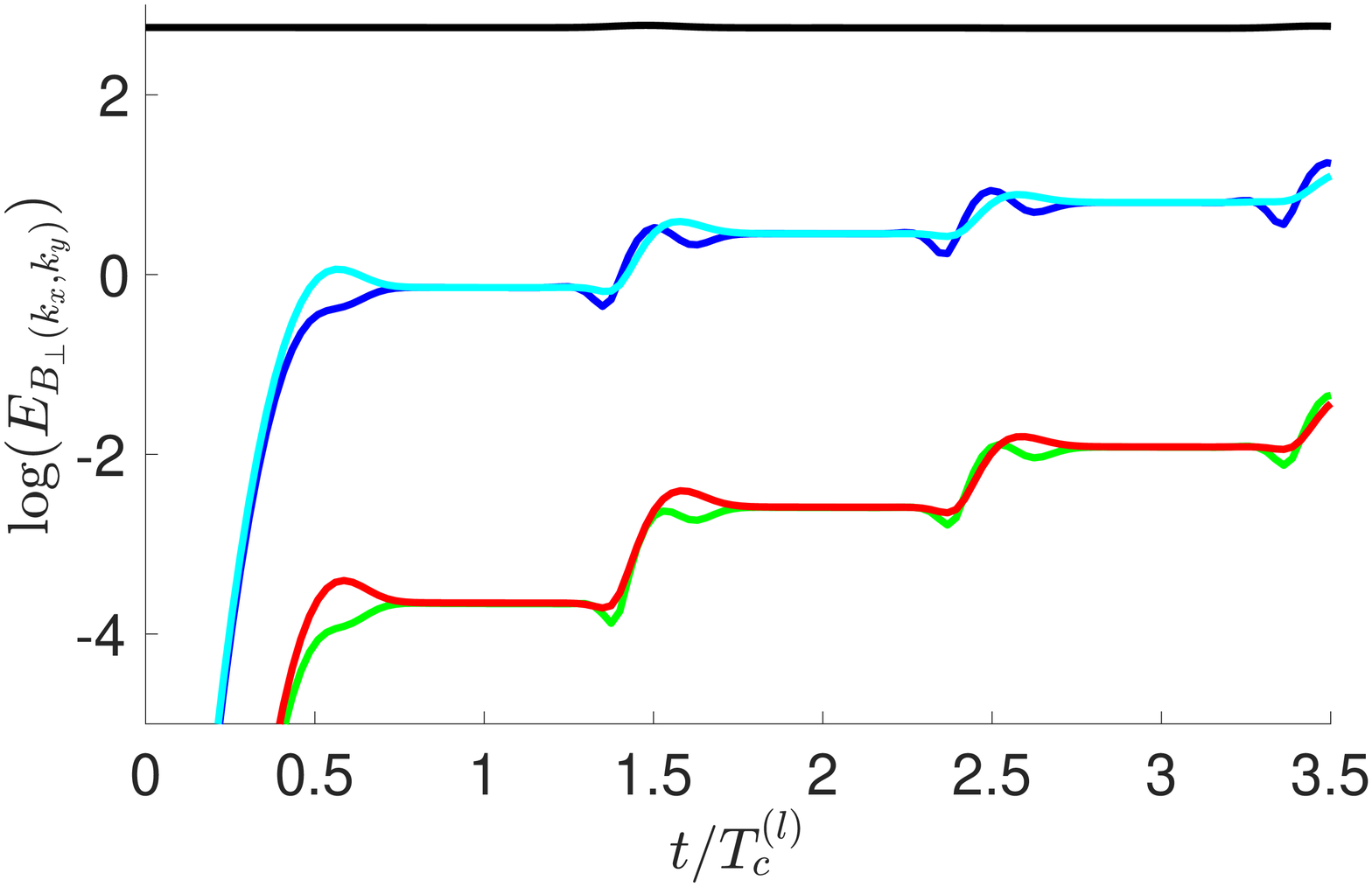}\hfill
 \includegraphics[scale = .25]{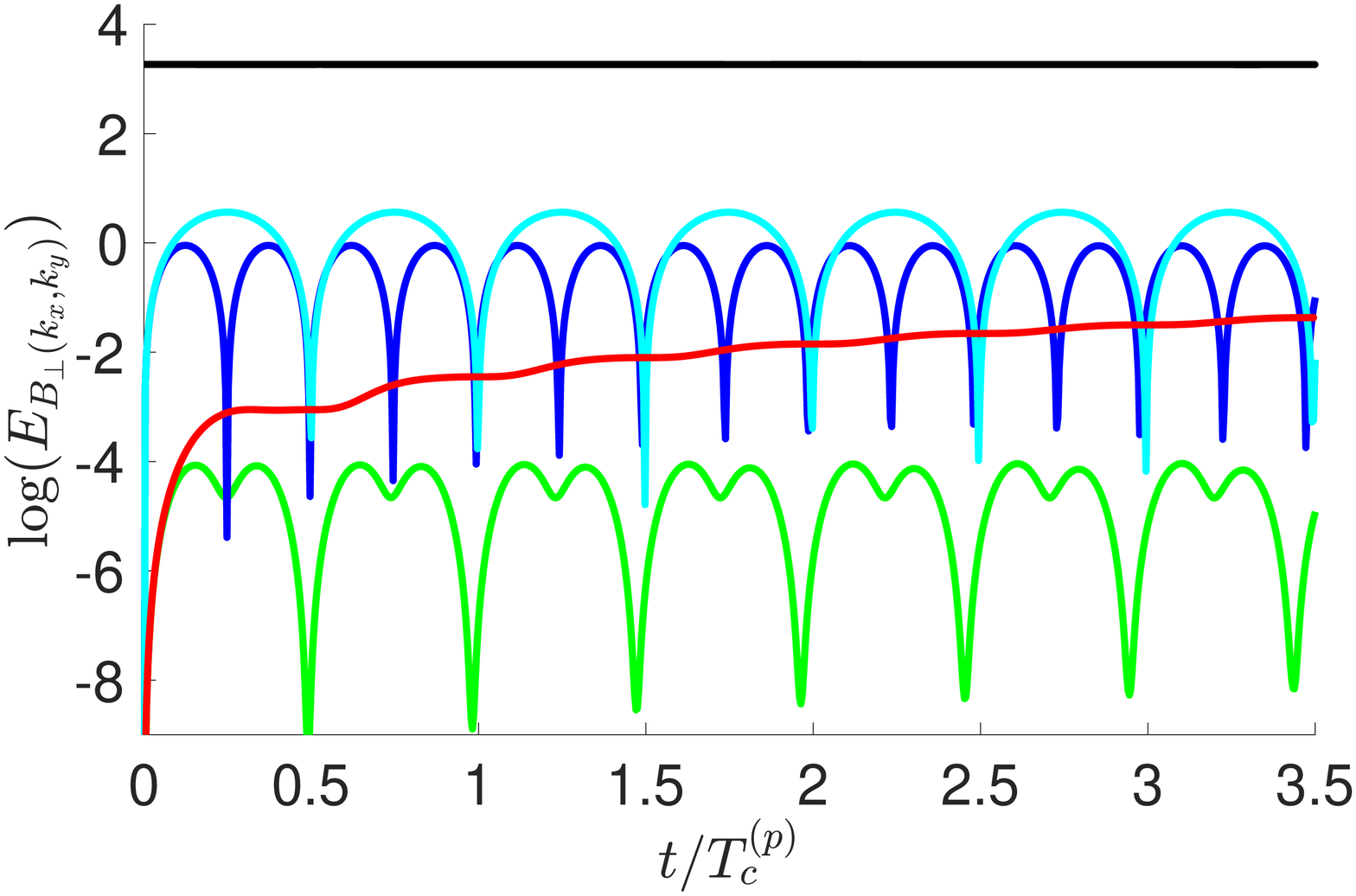}
 \vfill
 \hspace{0.1in} (e) Legend
 \vfill
 \hspace{0.1in} \includegraphics[scale = .4]{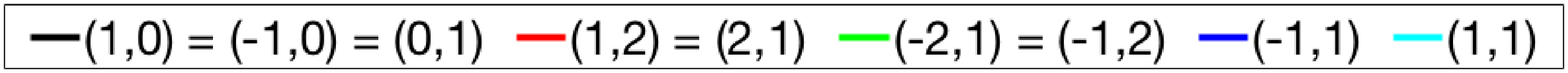}
 \caption{Energy evolution of each case for key $(k_x,k_y)$ modes after 3 collisions, for the (a) localized, strongly nonlinear case \T{LS}, (b) periodic strongly nonlinear case \T{PS}, (c) localized, weakly nonlinear case \T{LW}, (d) periodic, weakly nonlinear case \T{PW}.
 \label{fig:energy3col}}
 \end{figure}

\begin{figure}
 \hspace{0.1in} (a) \T{LS} \hspace{2.7 in}  (b) \T{PS}
 \vfill
 \includegraphics[scale = .25]{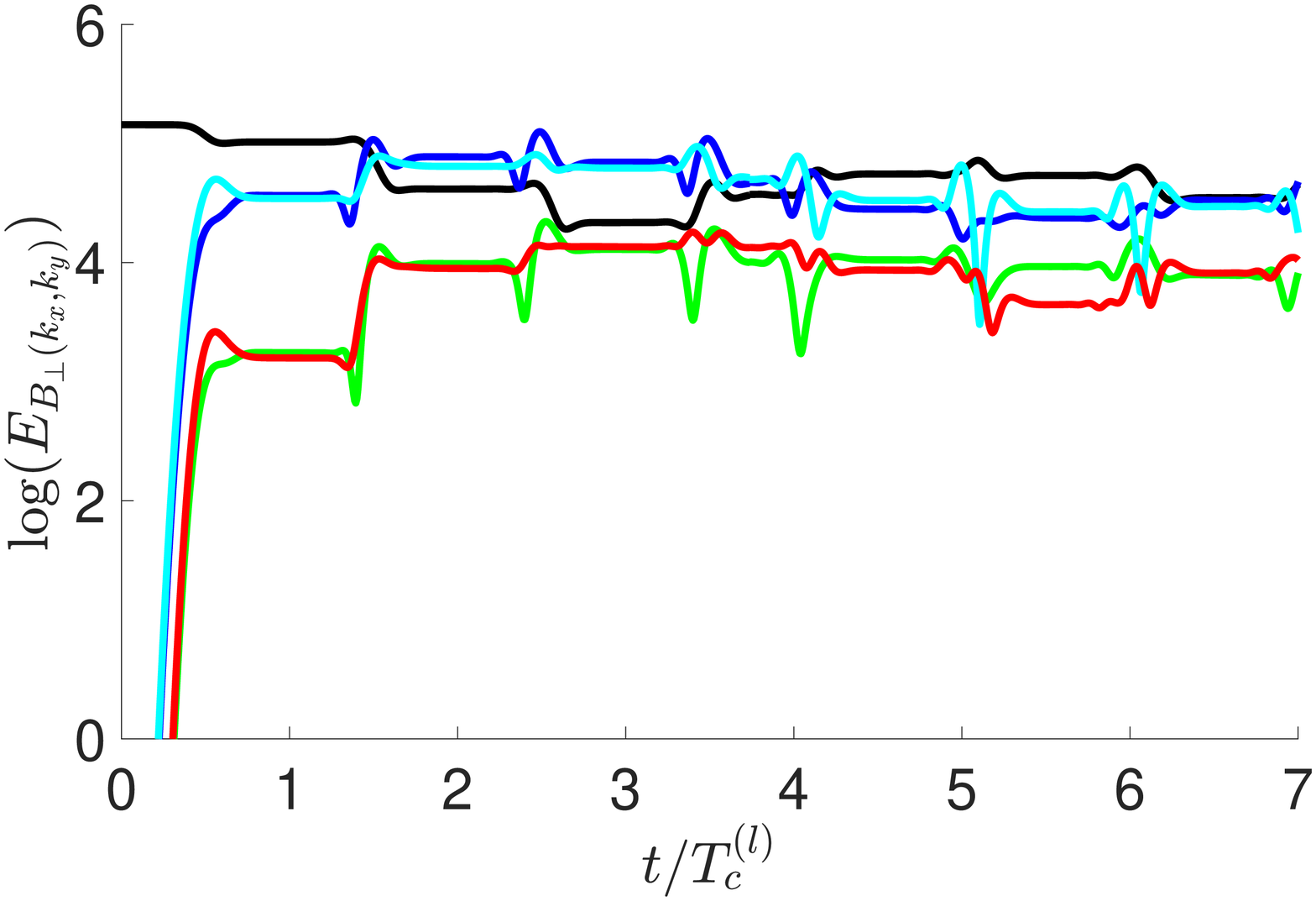}\hfill
 \includegraphics[scale = .25]{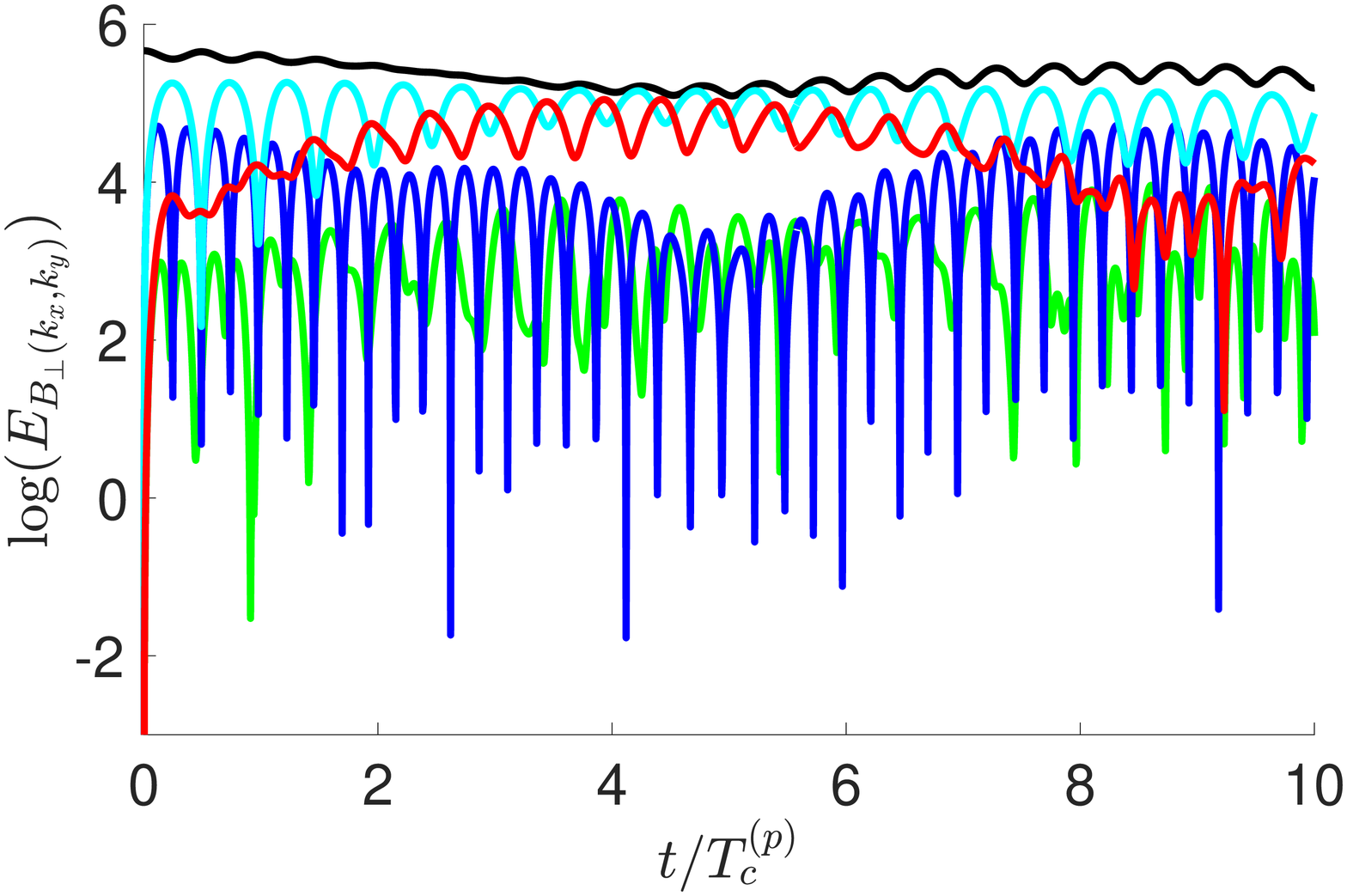}
 \vfill
 \hspace{0.1in} (c) \T{LW} \hspace{2.7 in}  (d) \T{PW}
 \vfill
 \includegraphics[scale = .25]{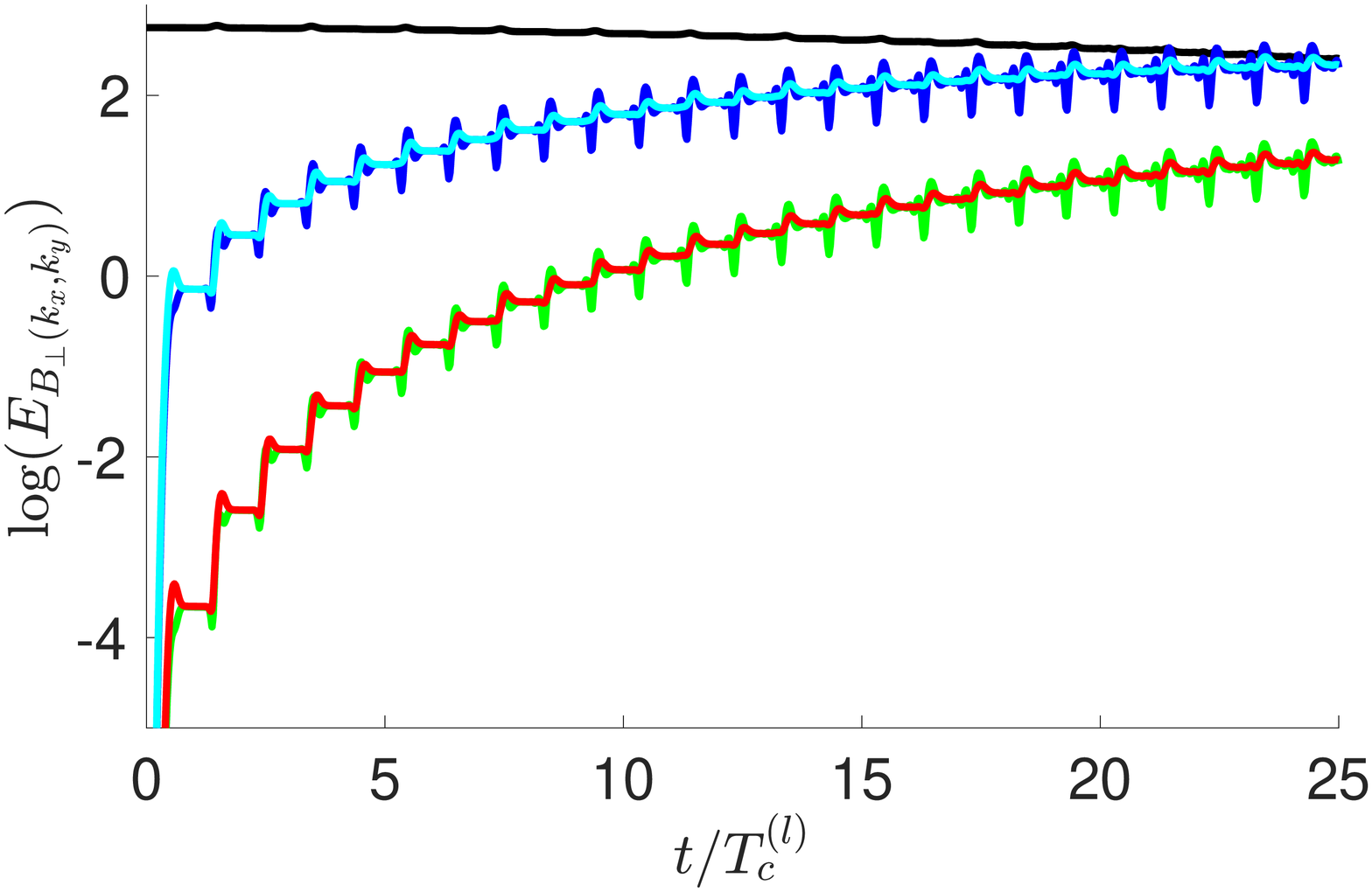}\hfill
 \includegraphics[scale = .25]{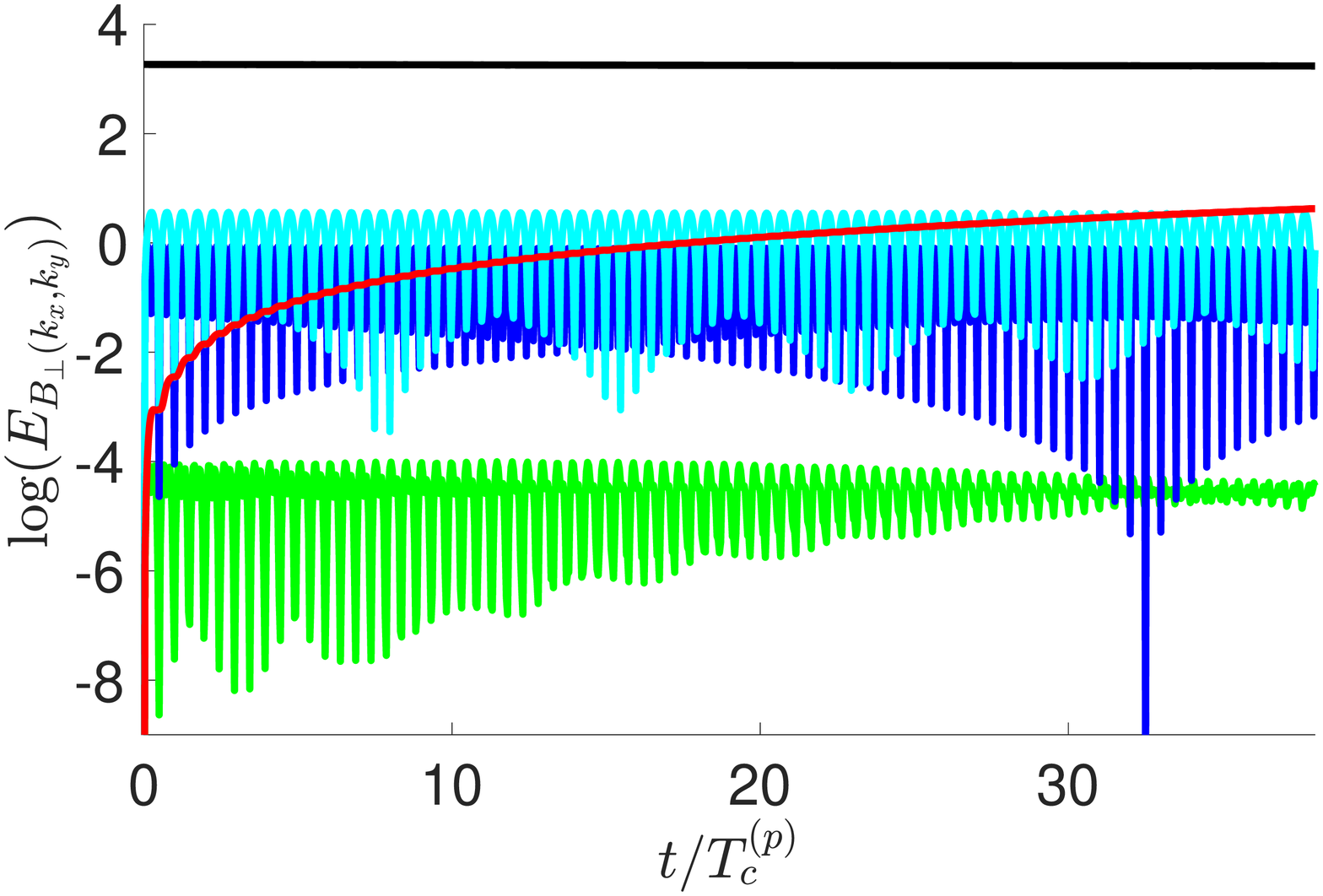}
 \vfill
  \hspace{0.1in} (e) Legend
 \vfill
 \hspace{0.1in} \includegraphics[scale = .4]{energy_legend.eps}
 \caption{Full energy evolution of each case for key $(k_x,k_y)$ modes, for the (a) localized, strongly nonlinear case \T{LS}, (b) periodic strongly nonlinear case \T{PS}, (c) localized, weakly nonlinear case \T{LW}, (d) periodic, weakly nonlinear case \T{PW}.
 \label{fig:fullenergy}}
 \end{figure}

 The temporal evolution of energy of select $(k_x,k_y)$ modes for the first three collisions is shown in \figref{fig:energy3col} while \figref{fig:fullenergy} shows the full time evolution of the simulations. To illustrate differences between the periodic and localized cases, we first focus on the weakly nonlinear limit. From panel (d), in the periodic case, the evolution agrees with the analytical solution from \citet{Howes:2013a}, as described qualitatively above in \secref{sec:intro}. Notice that the secondary (1,1) mode, which mediates the secular transfer of energy to the tertiary (1,2) and (2,1) modes, does not experience a net gain in energy. This $(1,1)$ mode corresponds to the inherently nonlinear fluctuation that does not propagate, as described in the introduction.  In contrast, the secondary (1,1) mode of the localized case in panel (c) clearly does gain energy, which is the most consequential difference among all the curves.  This means that in the localized case, this secondary mode gains energy like all other nonlinearly generated modes. One other major distinction between \T{LW} and \T{PW} is that in \T{LW}, energy is only transferred during periods when the wavepackets overlap in $z$, giving the energy evolution curve a stair-step appearance. In contrast, \T{PW} has persistent energy transfer since the wavepackets never separate. 
 
 Note that convergence studies have been done to verify that the results in this $(n_x,n_y) = (32,32)$ resolution are accurately resolved by the grid in \T{AstroGK}.  We initially started this experiment using a resolution of $(n_x,n_y)= (10,10)$ and replicated the same results using $(n_x,n_y) = (16,16)$.  For the $(n_x,n_y) = (32,32)$ case, we followed the evolution of energy until it deviated from the (16,16) resolution case and ceased the simulation at that point. The results presented in \figref{fig:fullenergy} follow the evolution of energy up until the time step of this deviation point for each of the localized and periodic cases in the weakly and strongly nonlinear limit. At the end of the time evolution shown in \figref{fig:fullenergy}, about 13\% of the initial magnetic energy has been transferred nonlinearly to higher $k_\perp$ modes (not shown in the figure) for case \T{LS} and about 17\% of the initial magnetic energy for case \T{PS}.

\subsection{Identification of Nonlinearly Generated Modes as \Alfven Waves}
\label{sec:ver}

In the periodic case, as reviewed in the introduction, the secondary (1,1) mode mediates the secular transfer of energy from the primary \Alfven waves to the tertiary \Alfven waves, and this mode is an inherently nonlinear fluctuation that satisfies neither the linear eigenfunction relation nor the linear dispersion relation for an \Alfven wave.  For the more realistic case of localized \Alfven wavepacket collisions, we aim to determine here the nature of the secondary (1,1) mode.  Specifically, we ask whether this secondary (1,1) mode is an \Alfven wave.  A linear \Alfven wave must satisfy two conditions \citep{Howes:2013a}: (i) it satisfies the linear eigenfunction relation for an \Alfven wave, $\mathbf{u}_\perp /v_A = \pm \delta  \mathbf{B}_\perp/ B_0$; and (ii) it has a frequency given by the linear \Alfven wave dispersion relation, $\omega = \pm k_\parallel v_A$.  The strongly nonlinear localized case \T{LS} is the most relevant to the case of heliospheric plasma turbulence, so we focus strictly on this case below.

\subsubsection{\Alfven Wave Eigenfunction Relation}

To confirm that the fluctuations that are nonlinearly generated by the \T{LS} \Alfven wave collisions have the character of linear \Alfven waves, we first verify that the electric and magnetic field fluctuations are related by the following linear eigenfunction relation \citep{Howes:2013a}

\begin{equation} \label{eq:eig}
\frac{\mathbf{B}_\perp}{B_0} = \pm \frac{c\mathbf{E}_\perp}{v_A B_0} \times \mathbf{\hat{z}}
\end{equation}
where the + sign corresponds to an \Alfven wave travelling down the magnetic field in the $-z$ direction, and the $-$ sign corresponds to an \Alfven wave travelling up the magnetic field in the $+z$ direction. 

Separating the two components perpendicular to the equilibrium magnetic field $\mathbf{B}_0 = B_0 \hat{\mathbf{z}}$ given by Equation \ref{eq:eig}, we note that \Alfven waves travelling up the magnetic field in the $+z$ direction will satisfy the relations
\begin{equation}
\frac{B_x}{B_0}= -\frac{c E_y}{v_A B_0}
\quad \quad \quad
\frac{B_y}{B_0}= +\frac{c E_x}{v_A B_0}
\end{equation}
and that \Alfven waves travelling down the magnetic field in the $-z$ direction will satisfy the relations
\begin{equation}
\frac{B_x}{B_0}= +\frac{c E_y}{v_A B_0}
\quad \quad \quad
\frac{B_y}{B_0}= -\frac{c E_x}{v_A B_0}
\end{equation}
For notational simplicity, we use a hat to denote these dimensionless magnetic and electric field components, 
$\hat{B}_j \equiv B_j/B_0$ and 
$\hat{E}_j \equiv c E_j/(v_A B_0)$.
Note that the propagation direction of the \Alfven wave is easily determined by computing the Poynting flux,
$\mathbf{S} = (c /4 \pi) \mathbf{E} \times \mathbf{B}$.

   \begin{figure}
   \hspace{0.03in} (a) Primary modes $t/T_c^{(l)}$=0  \hspace{.1 in}  (b) Primary modes $t/T_c^{(l)}$=1 \hspace{.1 in} (c) Primary modes $t/T_c^{(l)}$=2
\vfill
 {\includegraphics[scale=.22]{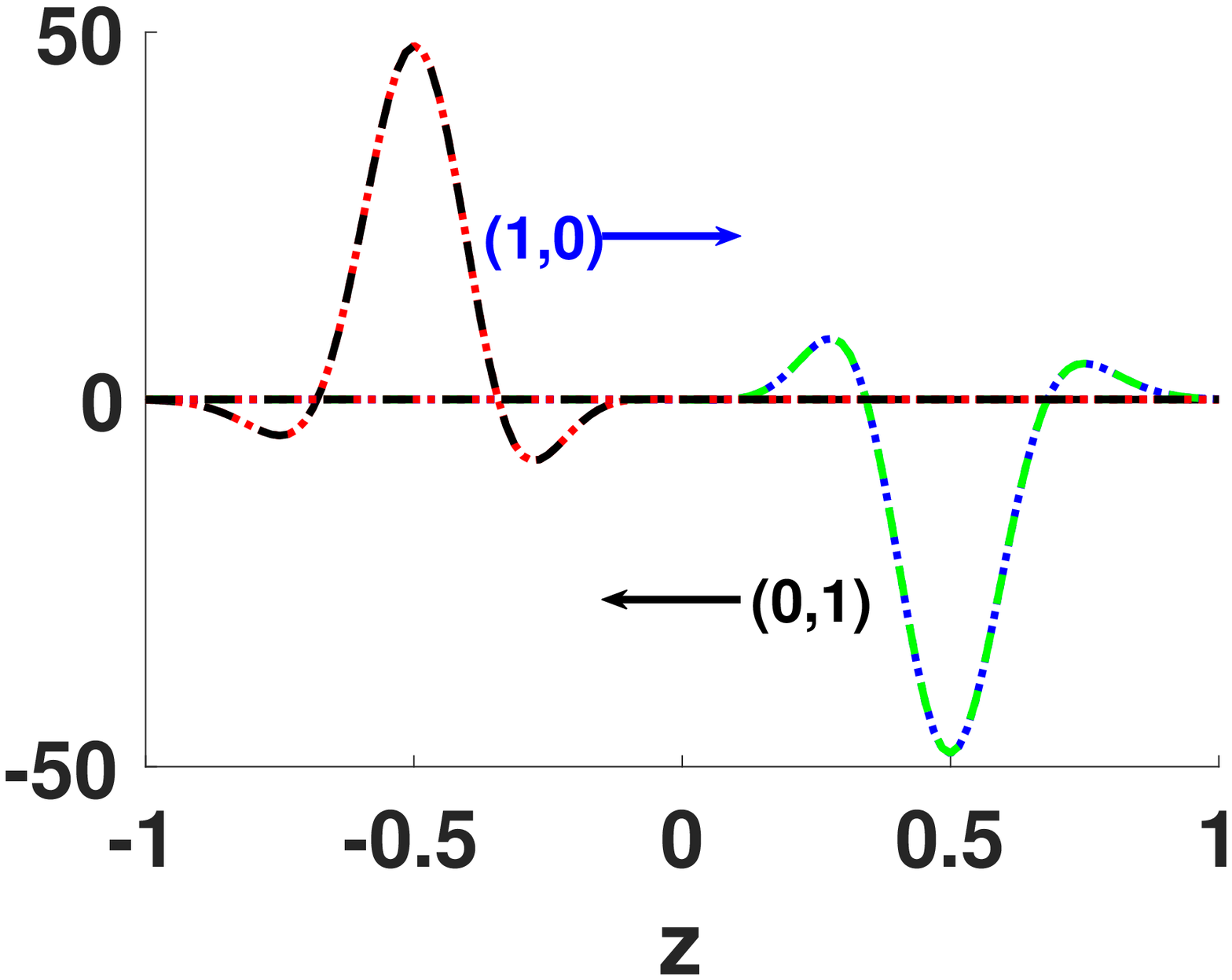}}	
 {\includegraphics[scale=.22]{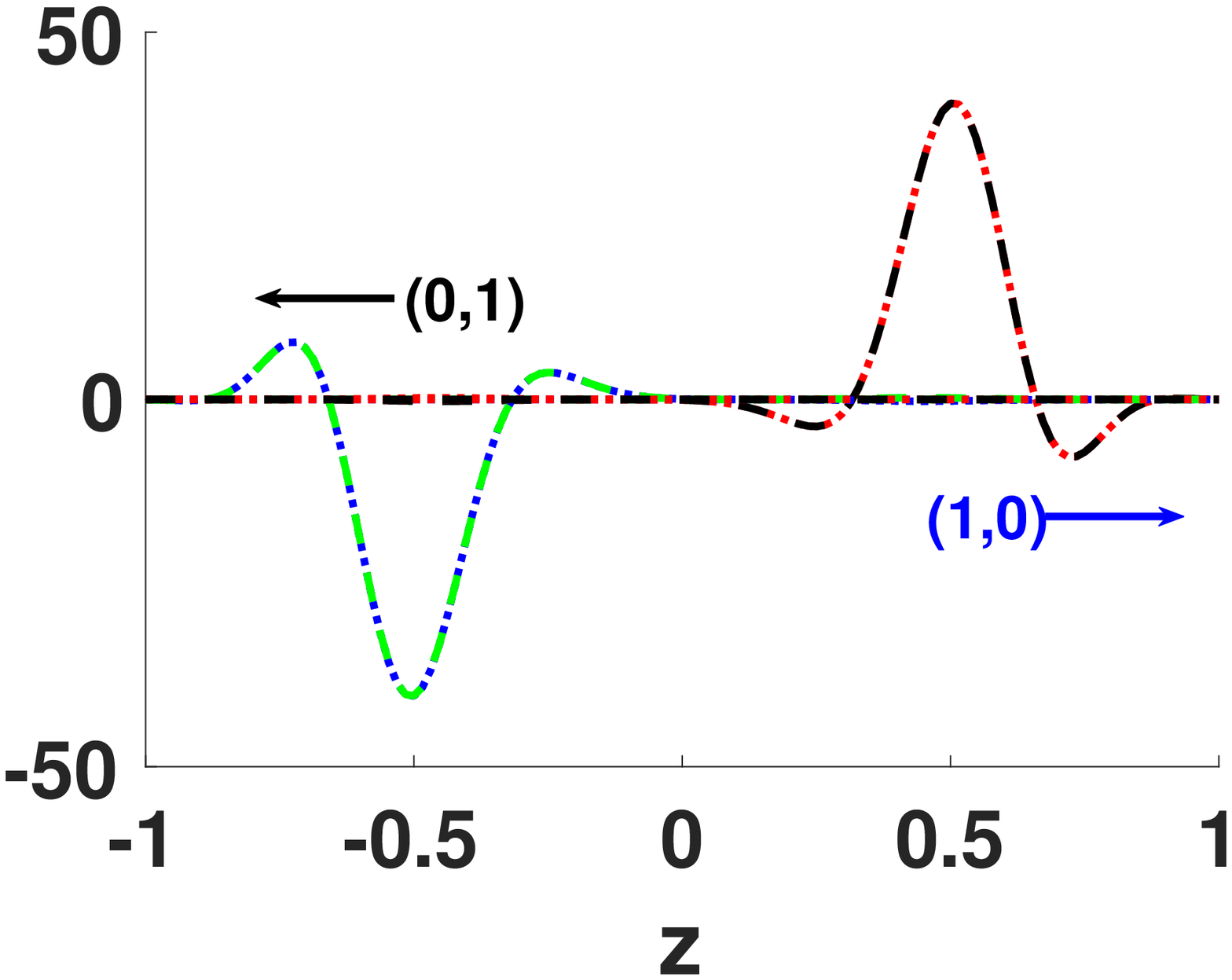}}
  {\includegraphics[scale=.22]{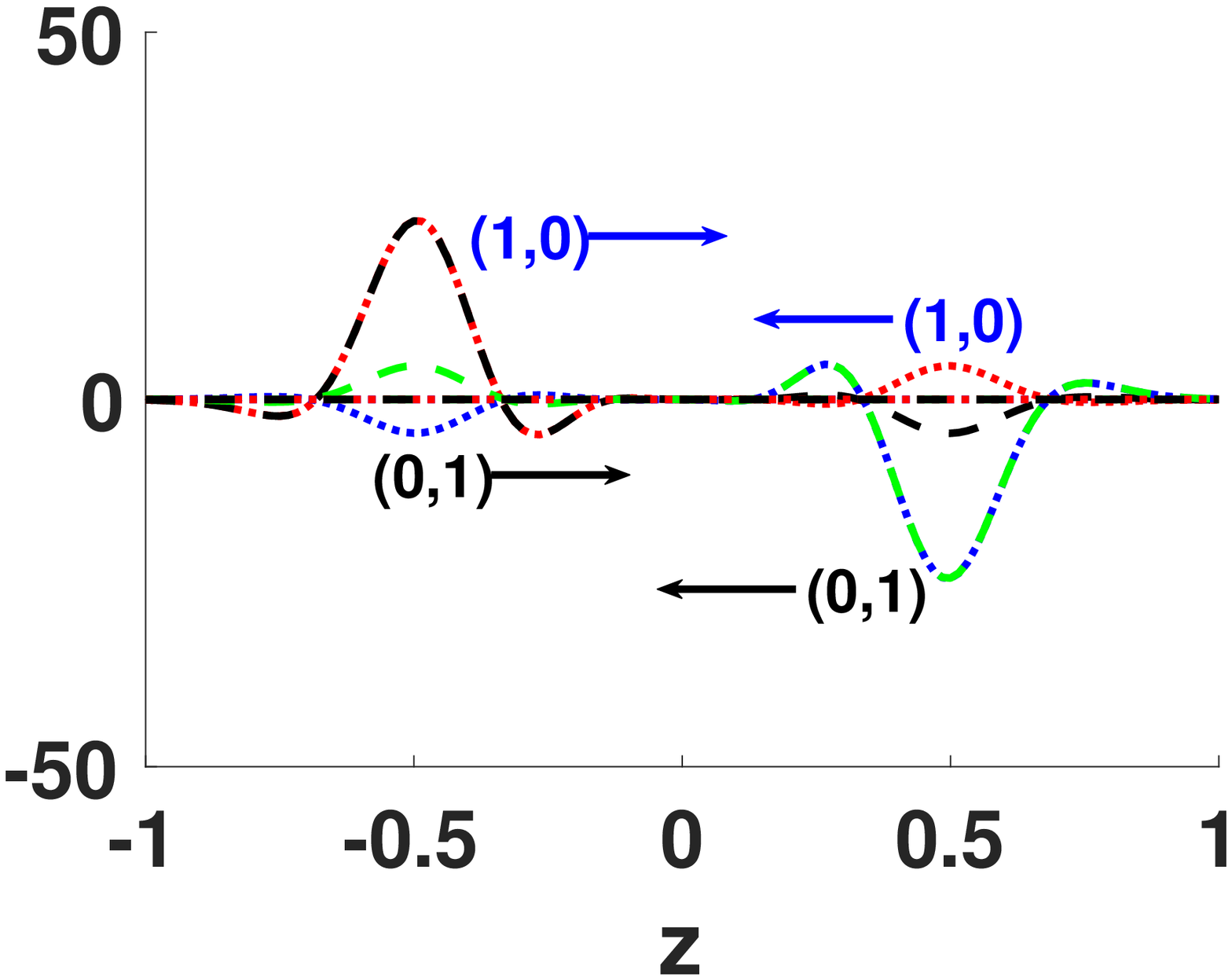}}
 \vfill
    \hspace{0.03in} (d) Secondary mode $t/T_c^{(l)}$=0  \hspace{.06 in}  (e) Secondary mode $t/T_c^{(l)}$=1 \hspace{.06 in} (f) Secondary mode $t/T_c^{(l)}$=2
\vfill
 {\includegraphics[scale=.22]{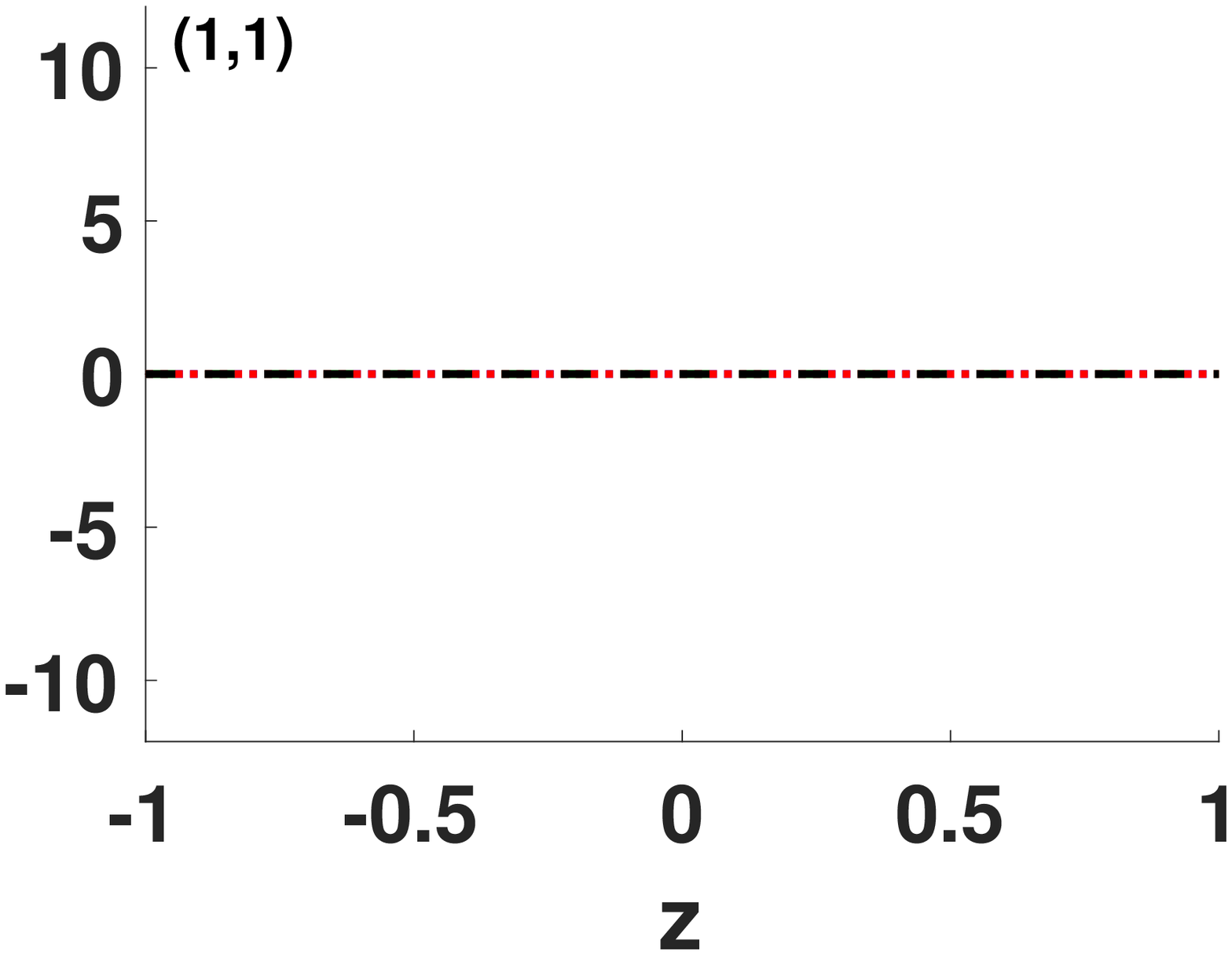}}
 {\includegraphics[scale=.22]{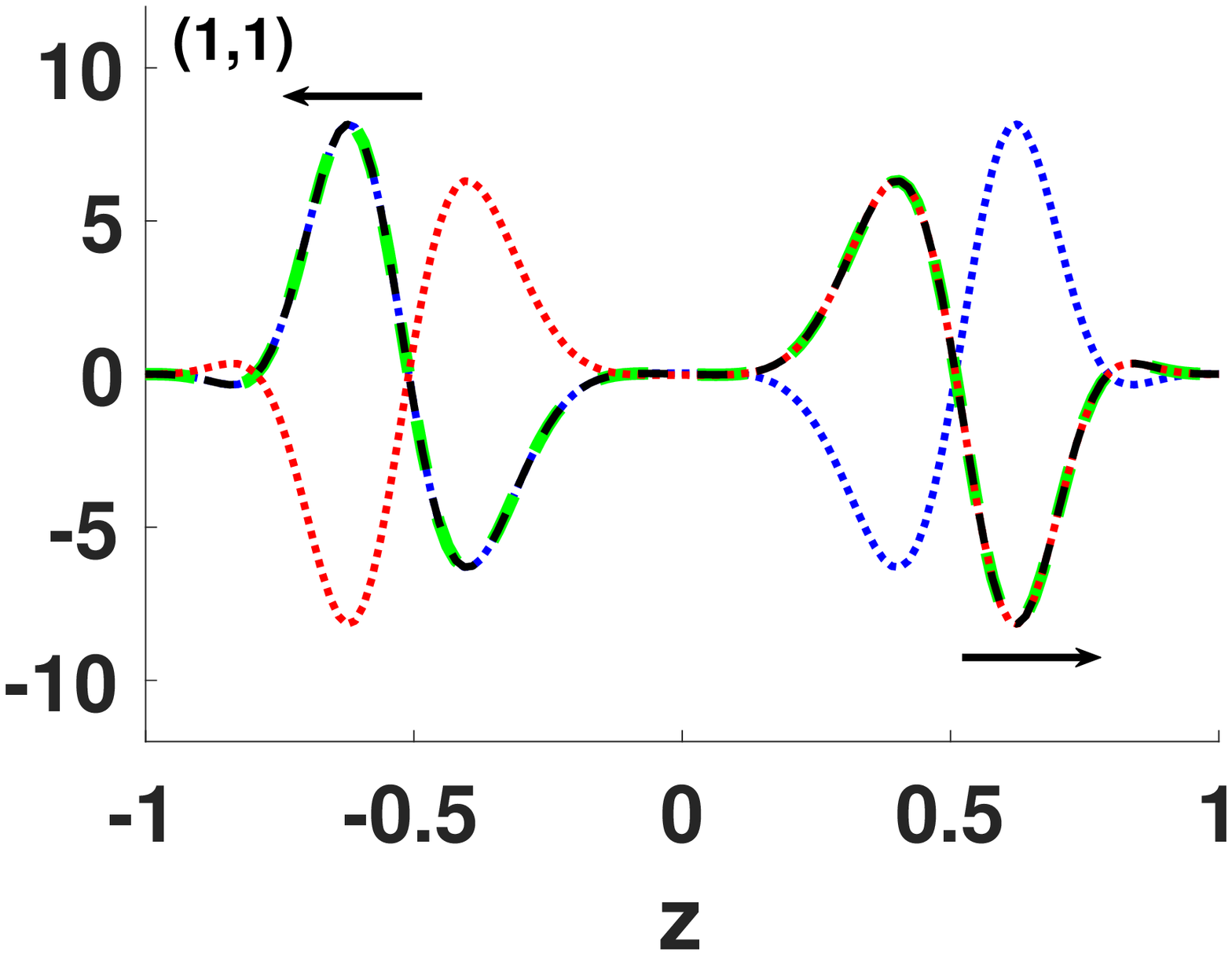}}
  {\includegraphics[scale=.22]{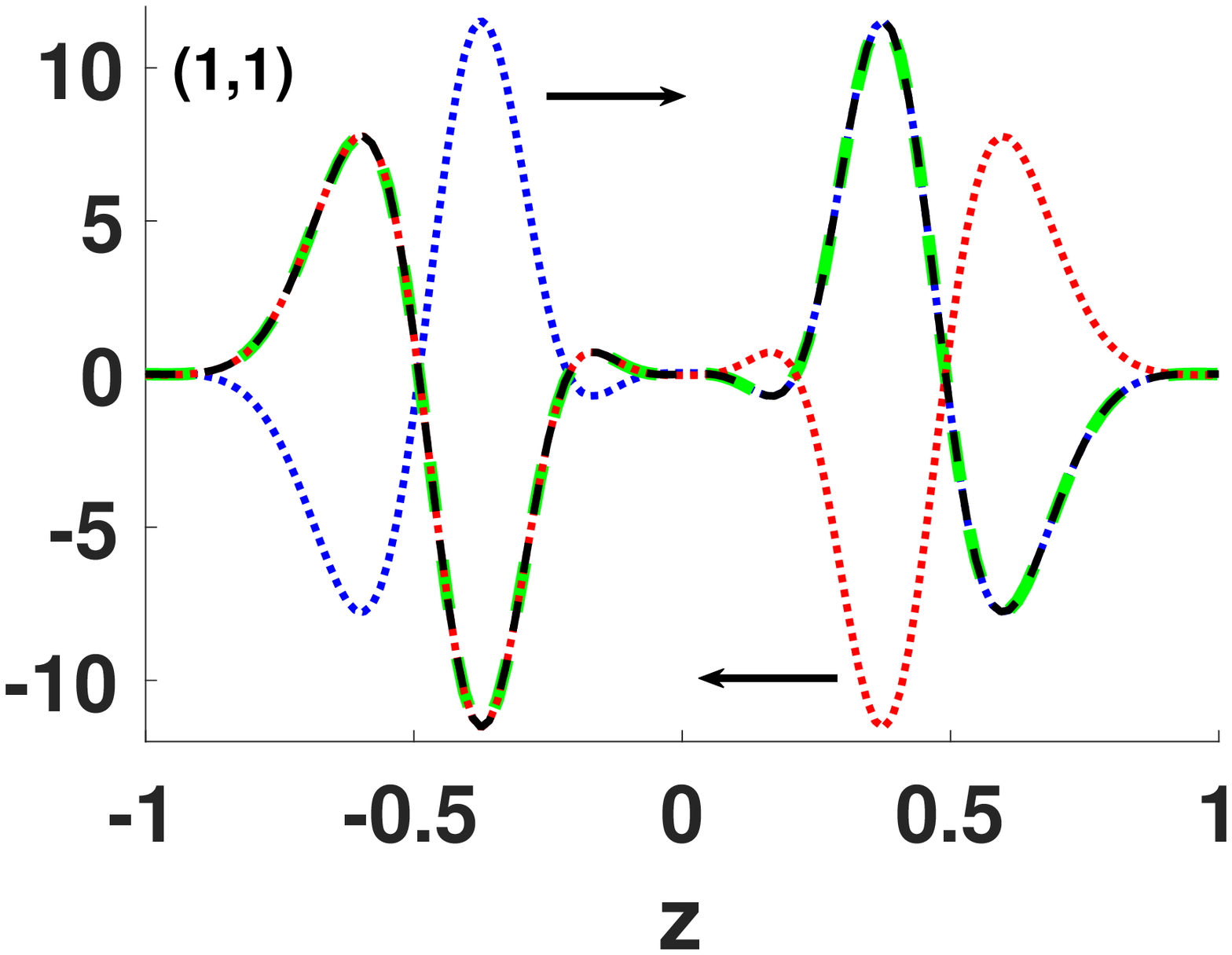}}
   \vfill
      \hspace{0.03in} (g) Tertiary mode $t/T_c^{(l)}$=0  \hspace{.2 in}  (h) Tertiary mode $t/T_c^{(l)}$=1 \hspace{.16 in} (i) Tertiary mode $t/T_c^{(l)}$=2
\vfill
 {\includegraphics[scale=.22]{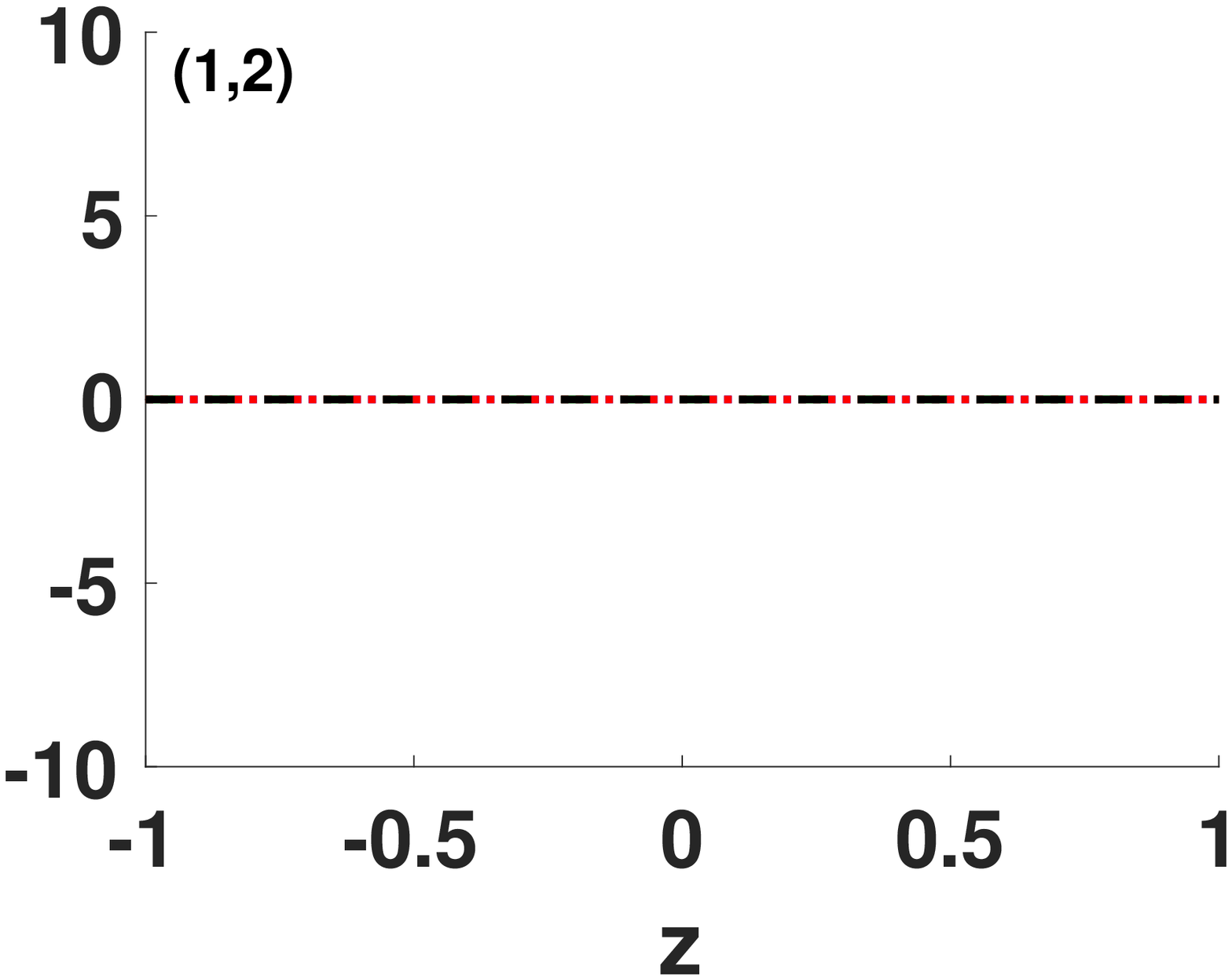}}
 {\includegraphics[scale=.22]{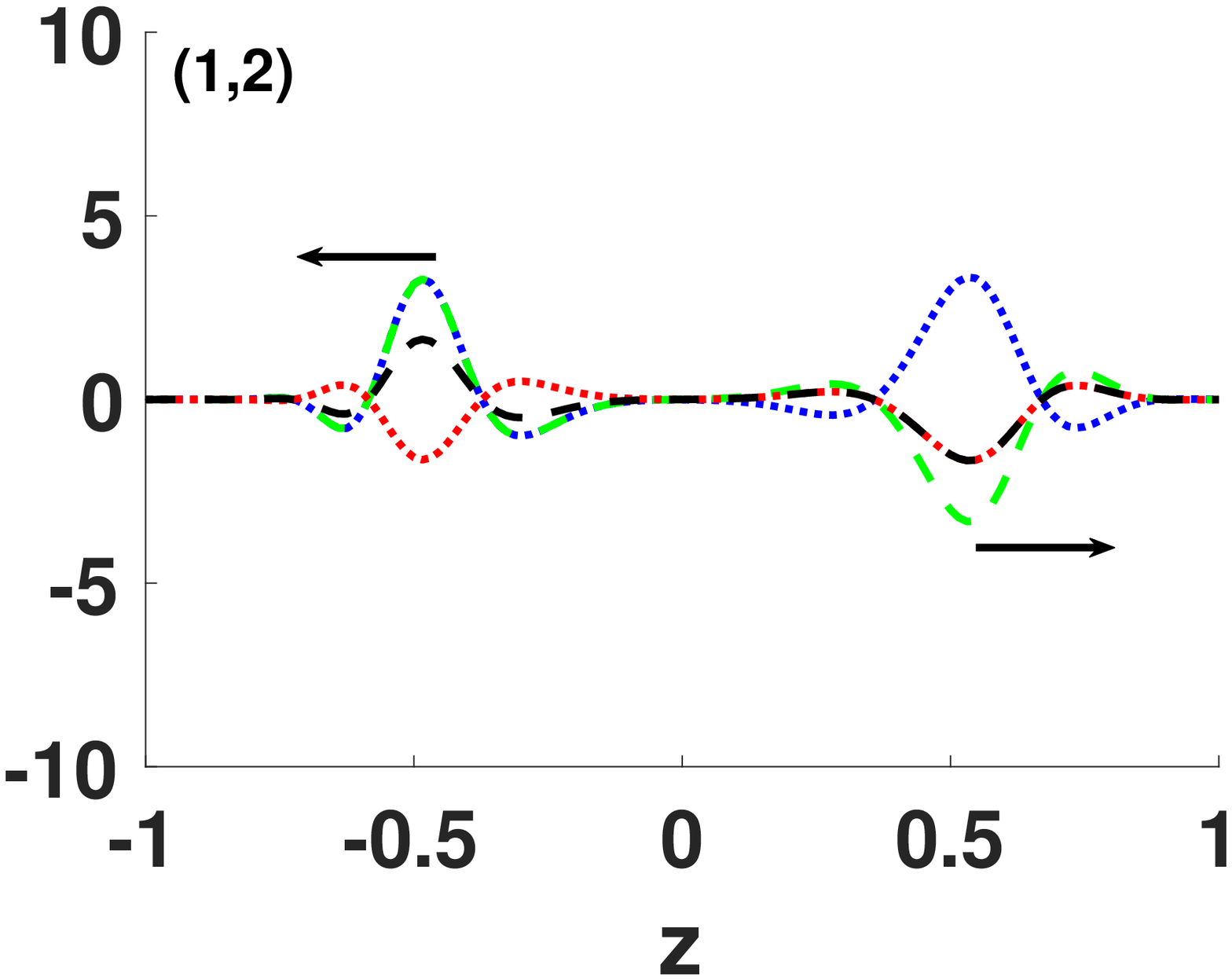}}
  {\includegraphics[scale=.22]{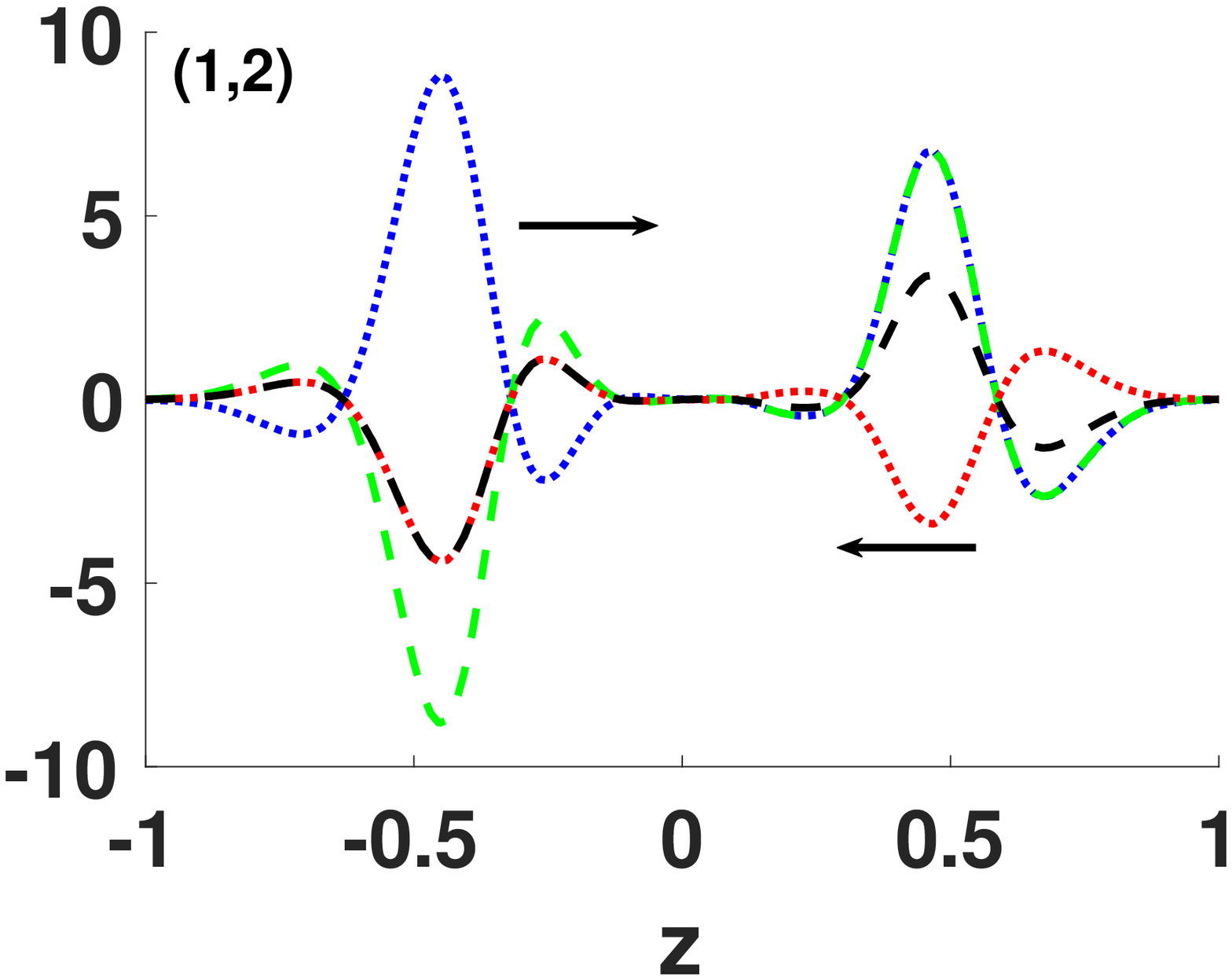}}
   \vfill
 \hspace{1.8in} (j) Legend
 \vfill
 \hspace{1.8in} \includegraphics[scale = .4]{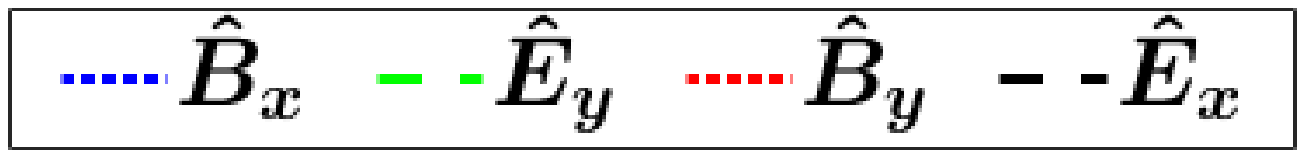}
 \caption{Snapshots in time of $\hat{B}_x$ (blue,dotted), $\hat{E}_y$ (green, dashed), $\hat{B}_y$ (red,dotted), and 
 $\hat{E}_x$ (black,dashed) of select $(k_x,k_y)$ Fourier modes in the \T{LS} case. The first, second, and third row corresponds to the primary, secondary, and tertiary modes respectively. All times are normalized to the localized \Alfven collision time, $T_c^{(l)}$. The black arrows indicate the direction of motion of the two colliding wavepackets.
 \label{fig:eig}}
 \end{figure}

 In \figref{fig:eig}, we present normalized $\hat{\mathbf{E}}$ and $\hat{\mathbf{B}}$ field components of the primary, secondary, and tertiary modes in the (first, second, and third row, respectively) at times $t/T_c^{(l)}=0, 1, 2$ (in the first, second, and third columns, respectively).  In the first column of \figref{fig:eig} at 
 $t/T_c^{(l)}=0$, we have only (a) the primary \Alfven wavepackets. The upward propagating \Alfven wave 
has a perpendicular variation given by the (1,0)  Fourier mode and has a magnetic field  polarization in the $y$ direction.  This wavepacket satisfies the normalized eigenfunction for an upward propagating \Alfven wave, $\hat{B}_y = \hat{E}_x$ (red/black).  The downward propagating \Alfven wave has a perpendicular variation given by the (0,1) Fourier mode and has a magnetic field polarization in the $x$ direction.  This wavepacket satisfies the normalized eigenfunction for a downward propagating \Alfven wave, $\hat{B}_x = \hat{E}_y$ (blue/green).  In addition, at $t/T_c^{(l)}=0$, (d) the secondary (1,1) Fourier mode and (g) the tertiary (1,2) Fourier mode are zero.

In the second column of \figref{fig:eig}, we show the primary, secondary, and tertiary modes after the first collision at $t/T_c^{(l)}=1$. In \figref{fig:eig}(b), the primary \Alfven waves have passed through each other completely and still satisfy the same linear \Alfven wave eigenfunction relations as before the first collision in (a). Shown in panel (e), energy has been transferred to the secondary (1,1) Fourier mode, in two separate localized wavepackets, each with magnetic field components in both the $x$ and $y$ direction.  At $z<0$, the downward propagating wavepacket satisfies the eigenfunction relations $\hat{B}_y = - \hat{E}_x$ (red/black) and $\hat{B}_x = \hat{E}_y$ (blue/green), as expected for a downward travelling \Alfven wave. At $z>0$, the upward propagating wavepacket satisfies the eigenfunction relations $\hat{B}_y = \hat{E}_x$ (red/black) and $\hat{B}_x = -\hat{E}_y$ (blue/green), as expected for an upward travelling \Alfven wave.  This confirms that this secondary (1,1) mode satisfies the linear \Alfven wave eigenfunction.  Shown in panel (h), the tertiary (1,2) Fourier mode also involves two separate localized wavepackets with magnetic field components in both the $x$ and $y$ direction.  A close inspection of the curves confirms that this tertiary (1,2) mode also satisfies the linear \Alfven wave eigenfunction. 

In the third column of \figref{fig:eig},  we show the primary, secondary, and tertiary modes after the second collision at $t/T^{(l)}_c=2$. In panel (c), the upward and downward moving Fourier wavepackets have developed a component of polarization perpendicular to their original polarizations.  For instance, the upward wavepacket, which initially (at $t/T_c^{(l)}=0$) consisted of only a (1,0) Fourier mode with magnetic field polarized in the $y$ direction (red), now has a smaller (0,1) Fourier mode contribution moving in the $+z$ direction that has a magnetic field polarized in the $x$ direction (blue).  Similarly, the downward moving wavepacket, originally solely involving a (0,1) Fourier mode polarized in the $x$ direction (blue), now also includes a smaller contribution from a (0,1) mode polarized in the $y$ direction (red).  These newly generated contributions to the upward and downward moving wavepackets gained energy through nonlinear energy transfer from other modes during the second collision.
The secondary and tertiary modes at $t/T^{(l)}_c=2$ in panels (f) and (i) also show an increase in amplitude relative to $t/T^{(l)}_c=1$, showing that nonlinear interactions in the second collision have further transferred energy to those modes from the primary \Alfven wavepackets.

\begin{figure}
      \hspace{0.03in} (a) Primary modes $t/T_c^{(l)}$=0  \hspace{.1 in}  (b) Primary modes $t/T_c^{(l)}$=.75 \hspace{.1 in} (c) Primary modes $t/T_c^{(l)}$=2
\vfill
   
 {\includegraphics[scale = .22]{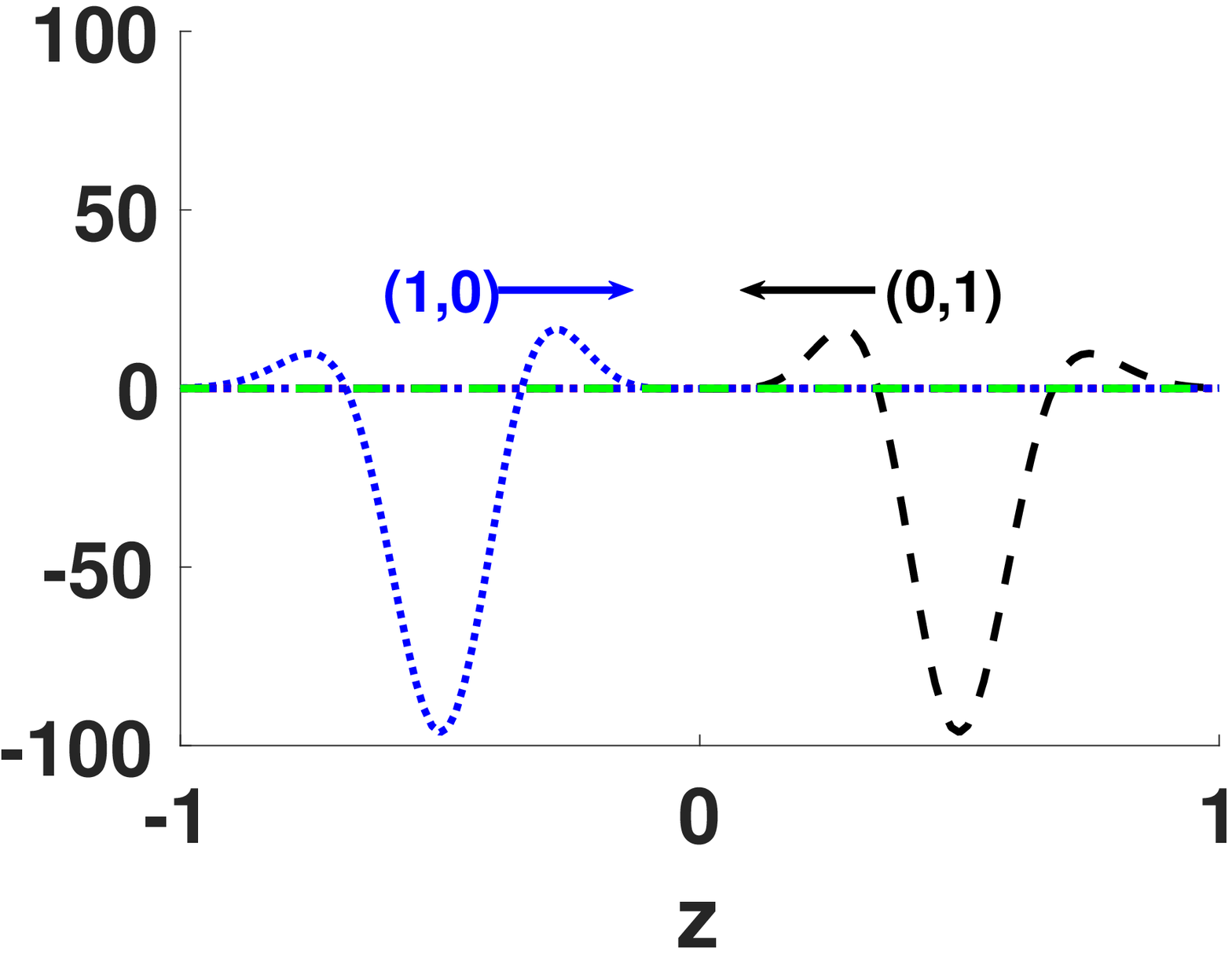}}	
 {\includegraphics[scale = .22]{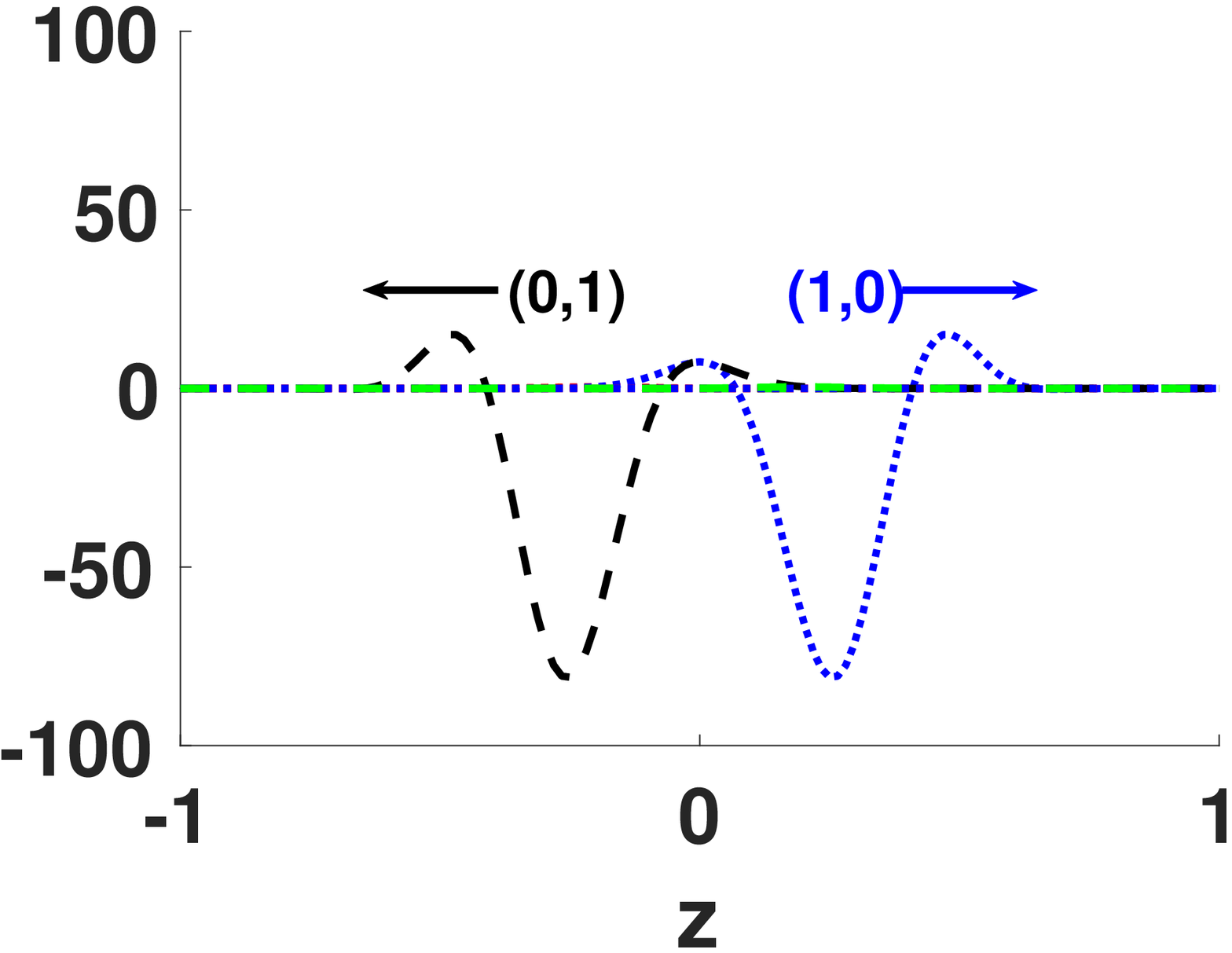}}
  {\includegraphics[scale = .22]{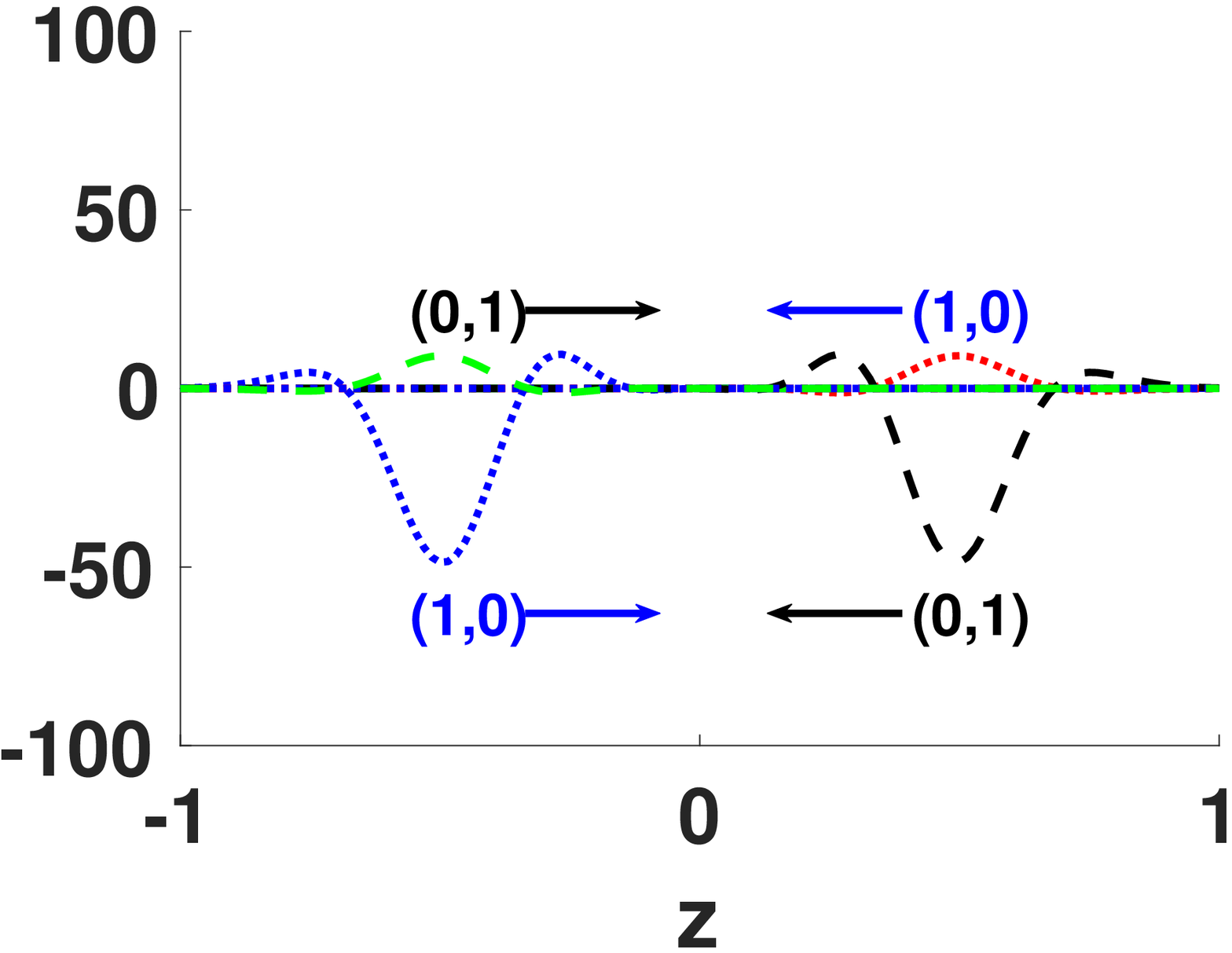}}
 \vfill
    \hspace{0.03in} (d) Secondary mode $t/T_c^{(l)}$=0  \hspace{.06 in}  (e) Secondary mode $t/T_c^{(l)}$=.75 \hspace{.07 in} (f) Secondary mode $t/T_c^{(l)}$=2
\vfill
 {\includegraphics[scale = .22]{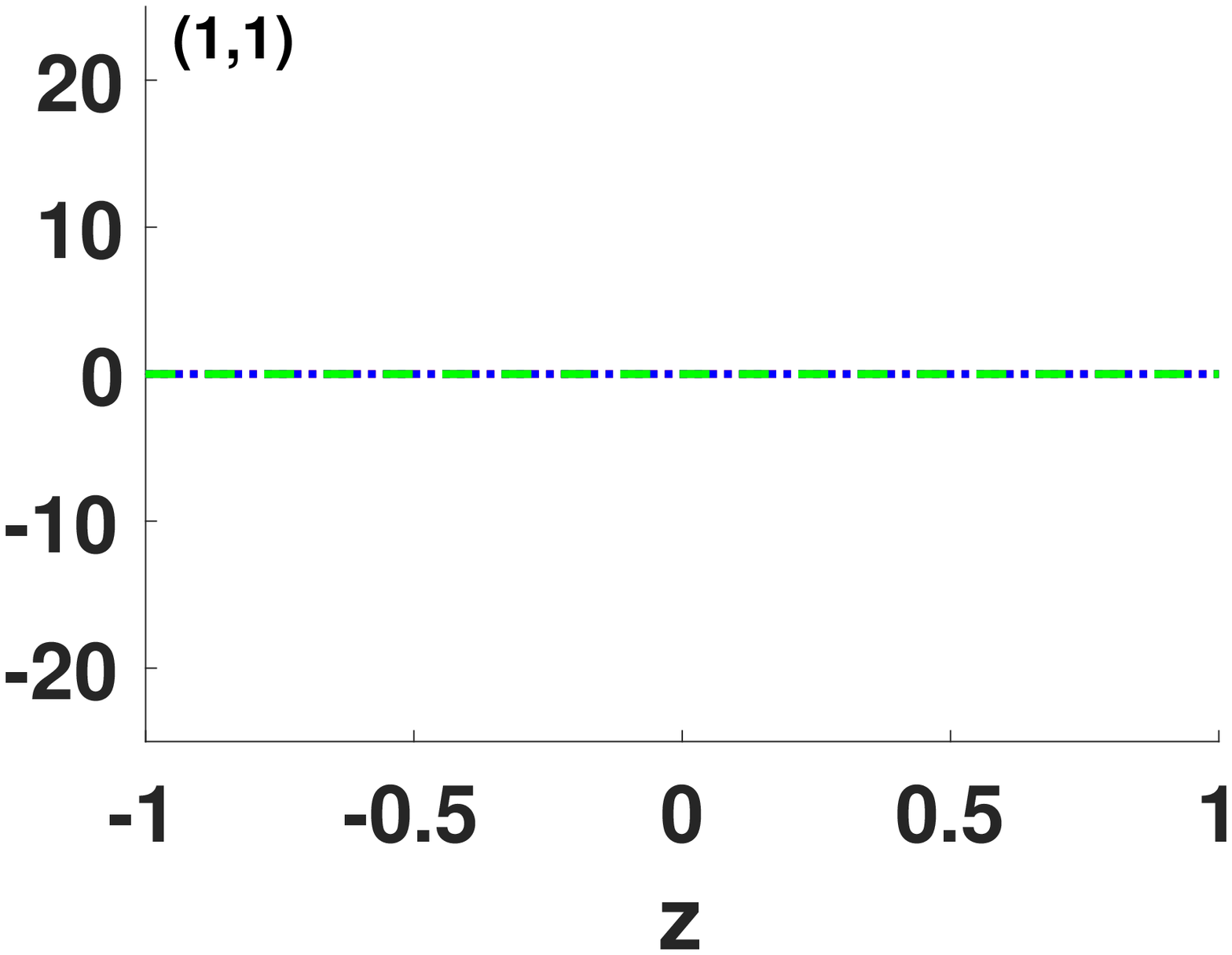}}
 {\includegraphics[scale = .22]{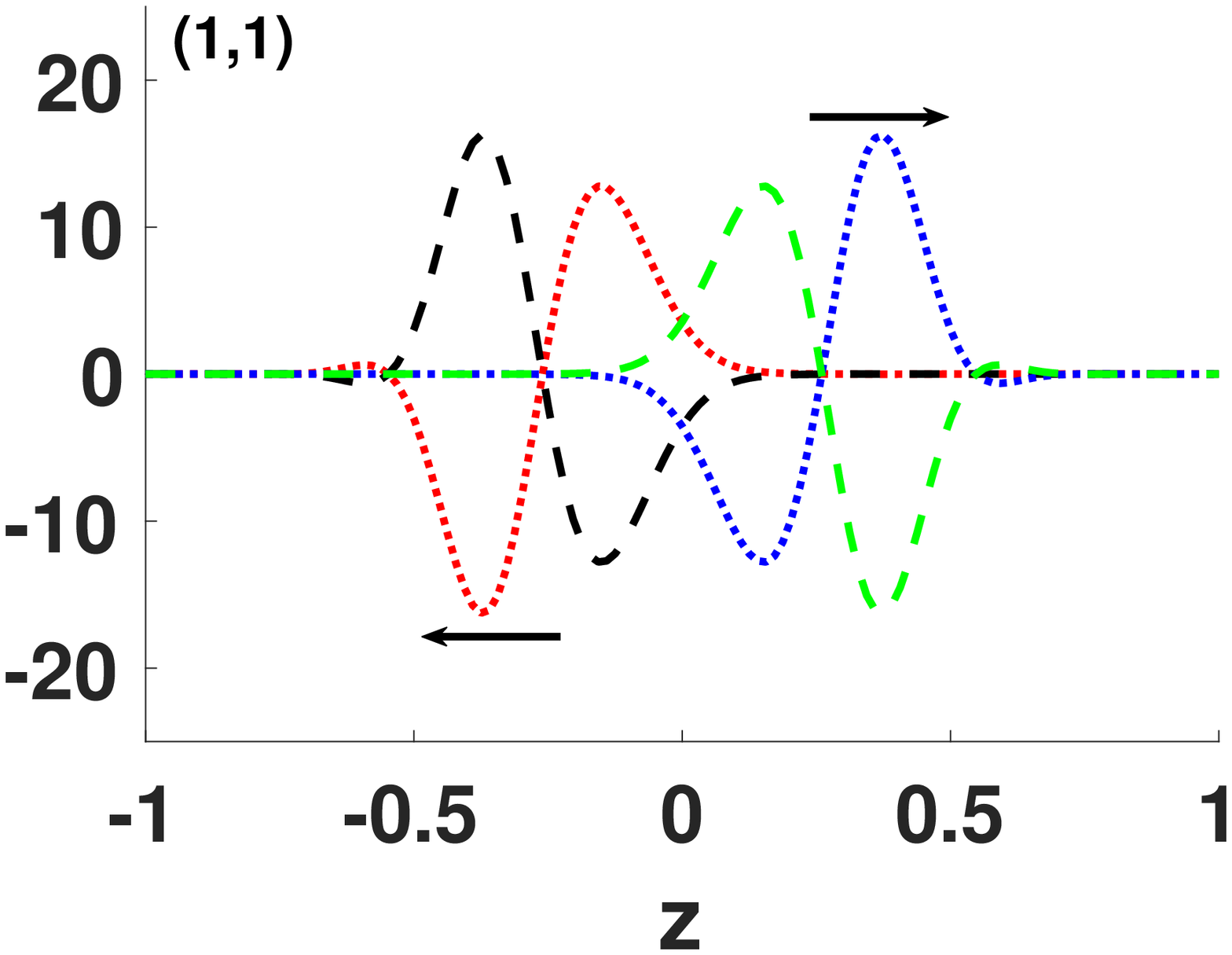}}
  {\includegraphics[scale = .22]{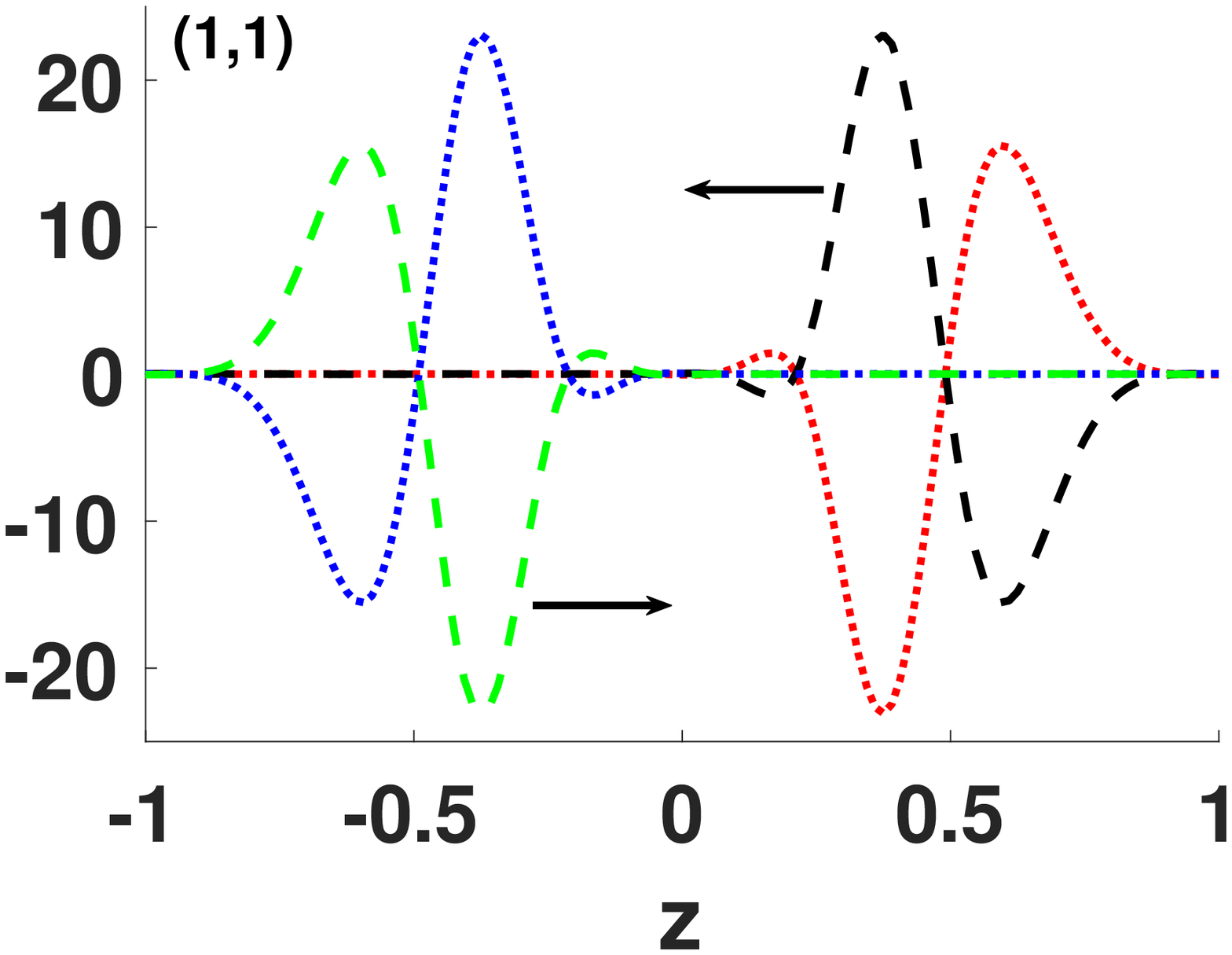}}
   \vfill
         \hspace{0.03in} (g) Tertiary mode $t/T_c^{(l)}$=0  \hspace{.2 in}  (h) Tertiary mode $t/T_c^{(l)}$=.75 \hspace{.16 in} (i) Tertiary mode $t/T_c^{(l)}$=2
\vfill
 {\includegraphics[scale = .22]{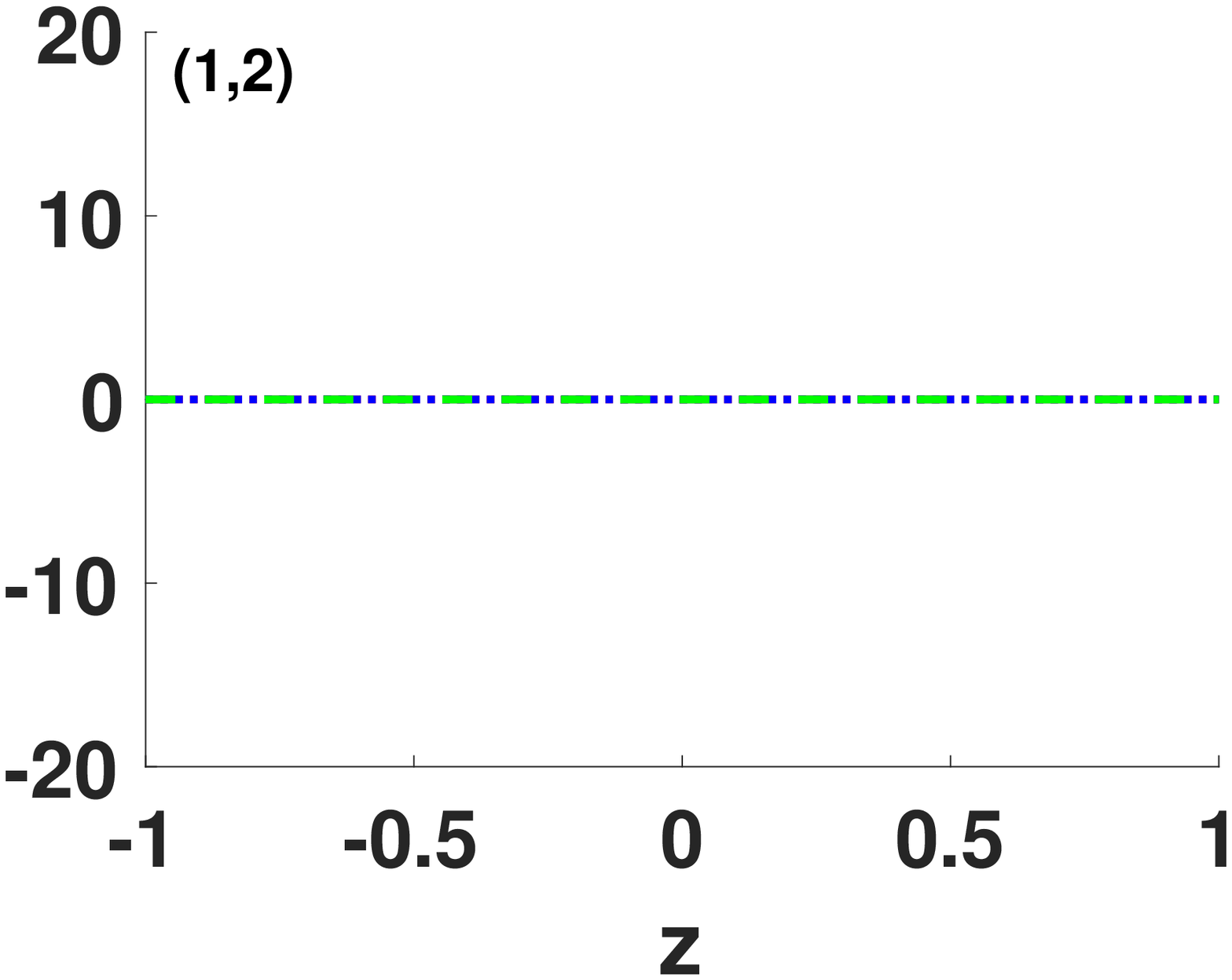}}
 {\includegraphics[scale = .22]{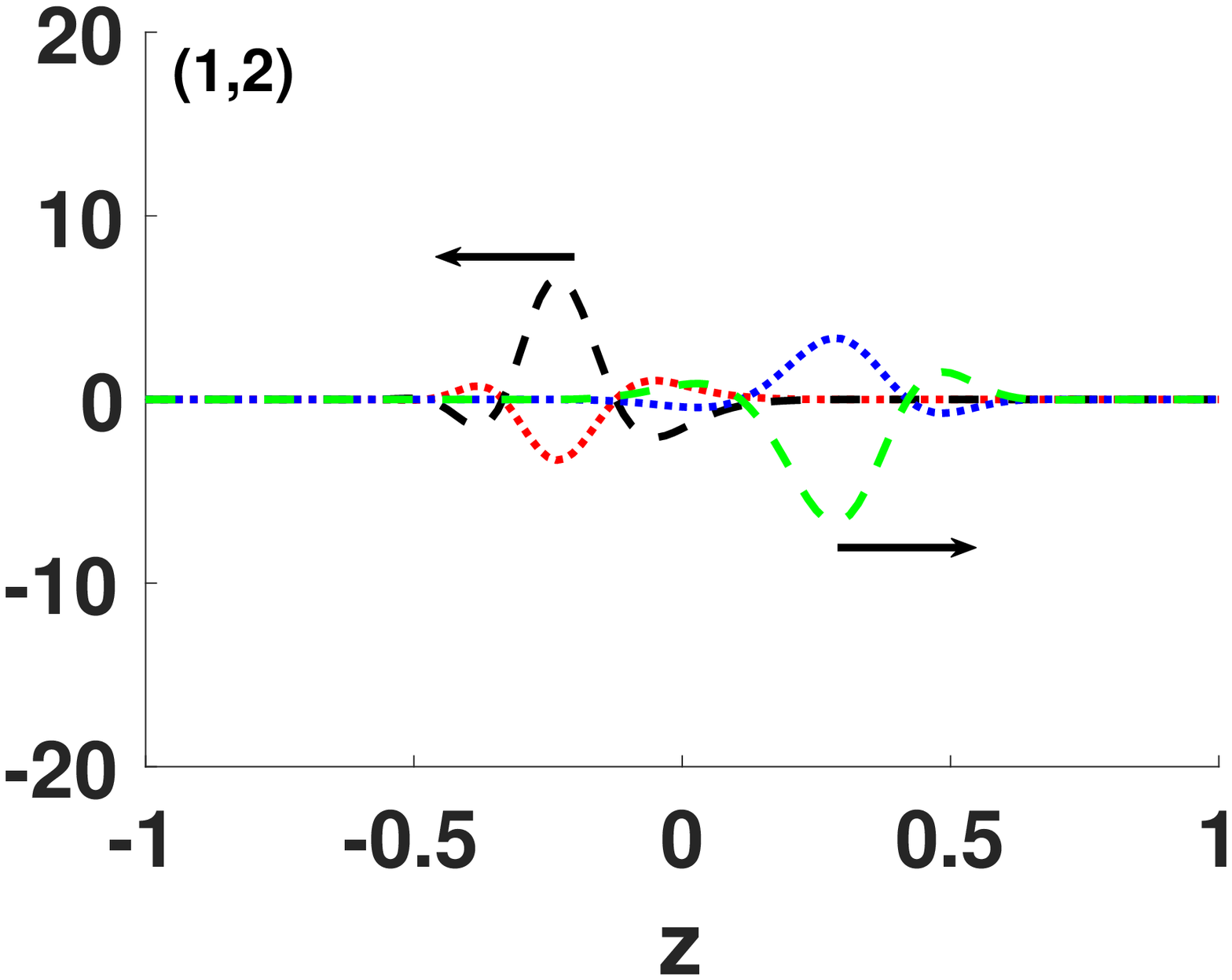}}
  {\includegraphics[scale = .22]{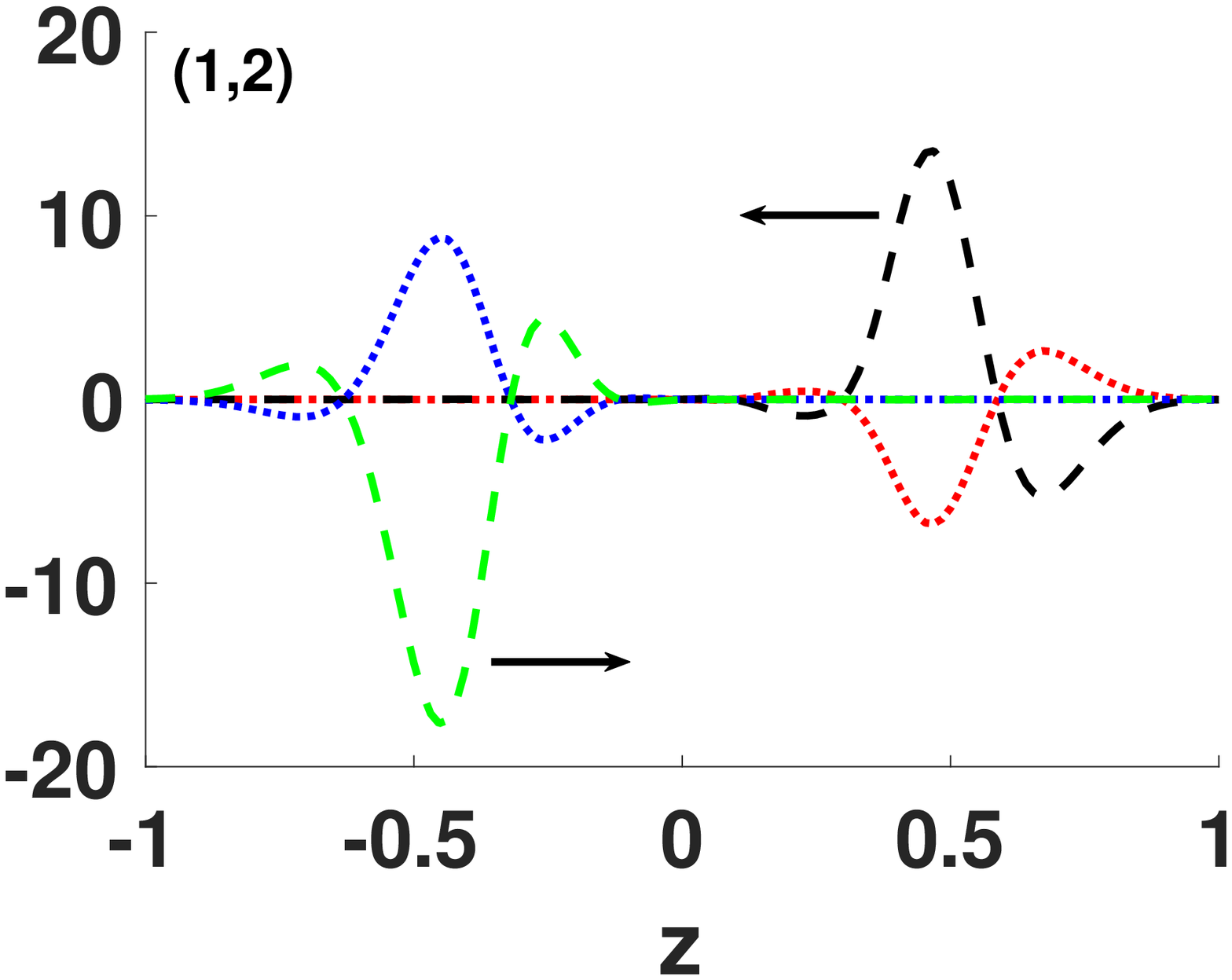}}
   \vfill
 \hspace{1.8in} (j) Legend
 \vfill
 \hspace{1.8in} \includegraphics[scale = .4]{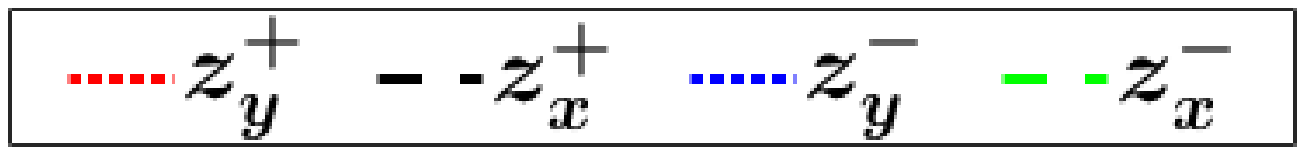}
 \caption{Snapshots in time of perpendicular Elssaser field components $z^+_y$ (red,dotted), $z^-_x$ (green, dashed), $z^-_y$ (blue,dotted), and $z^+_x$ (black,dashed) of key $(k_x,k_y)$ modes in the \T{LS} case. The first, second, and third row corresponds to the primary, secondary, and tertiary modes respectively. All times are normalized to the localized \Alfven collision time, $T_c^{(l)}$. The black arrows indicate the direction of motion of the two colliding wavepackets.
 \label{fig:els}}
 \end{figure}

Another way to visualize the upward and downward propagating \Alfven waves is to compute the Elssaser fields, $\mathbf{z}^\pm$. Specifically, we write the components of the normalized Elssaser variables for the upward ($z^-$) and downward ($z^+$) \Alfven waves as
\begin{equation}
\hat{z}_x^\pm \equiv \frac{z_x^\pm }{v_A} = \frac{c E_y}{v_A B_0}  \pm  \frac{\delta B_x}{B_0} 
\end{equation}
and 
\begin{equation}
\hat{z}_y^\pm \equiv \frac{z_y^\pm }{v_A} = -\frac{c E_x}{v_A B_0}  \pm  \frac{\delta B_y}{B_0} .
\end{equation}
In \figref{fig:els}, we plot the downward travelling Elsasser components $z_x^+$ (black) and $z_y^+$ (red) and the upward travelling Elsasser components $z_x^-$ (green) and $z_y^-$ (blue) for the same primary (first row), secondary (second row), and tertiary (third row) modes shown in \figref{fig:eig}. Note that in each of the two separate, counterpropagating wavepackets, the downward moving components (red/black) are always together in the same wavepacket localized in $z$, and likewise the upward moving components (blue/green) are always together, confirming the fact that these wavepackets remain localized in their extent along the equilibrium magnetic field.
 
The main message from \figref{fig:eig} and \figref{fig:els} is that, in the localized, strongly nonlinear \Alfven wavepacket collision (\T{LS}) case, all of the nonlinearly generated components of the \Alfven wavepackets satisfy the linear \Alfven wave eigenfunction condition given by \eqref{eq:eig}. This includes the secondary (1,1) Fourier mode, which does not satisfy this eigenfunction condition in the periodic case \citep{Howes:2013a,Nielson:2013a}.  Note that this characteristic of the difference between the periodic \Alfven wave and localized \Alfven wavepacket collisions is true in both the weakly and strongly nonlinear limits (not shown).

\subsubsection{\Alfven Wave Dispersion Relation}

In the MHD limit $k_\perp \rho_i \ll 1$, the \Alfven wave satisfies the linear dispersion relation $\omega = |k_\parallel| v_A$, where we adopt the convention that $\omega \ge 0$, so the sign of $k_\parallel$ indicates the direction of propagation of a plane \Alfven wave along the equilibrium magnetic field, $\V{B}_0 = B_0 \zhat$. This simple dispersion relation indicates that \Alfven waves are non-dispersive.  The parallel phase velocity is given by $v_{p_\parallel} = \omega/k_\parallel = \pm v_A$ and indicates that wave crests of constant phase propagate up or down the equilibrium magnetic field at the \Alfven speed, $v_A$.  The parallel group velocity is given by $v_{g_ \parallel} = \partial \omega/\partial k_\parallel = \pm v_A$, meaning that the envelope of an \Alfven wavepacket will propagate up or down the equilibrium magnetic field at the \Alfven speed, $v_A$.  

\begin{figure}
          \hspace{1.4 in} (a) Primary and secondary modes $t/T_c^{(l)}$=1 
\vfill
\centering {\includegraphics[trim=0 80bp 0 80bp,clip,scale=.5]{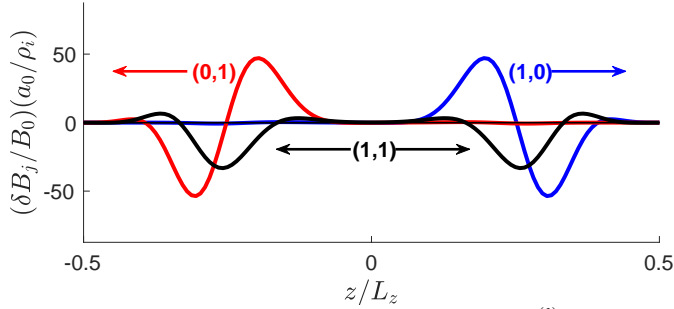}}
\vfill
          (b) Primary and secondary modes $t/T_c^{(l)}$=2 
\vfill
 \centering{\includegraphics[trim=0 80bp 0 80bp,clip,scale=.5]{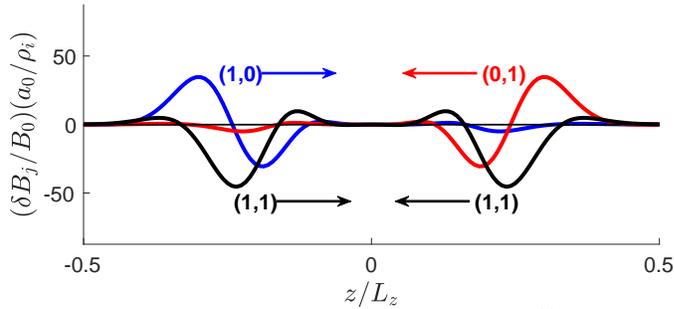}}
 \vfill
            (c) Primary and secondary modes $t/T_c^{(l)}$=3 
\vfill
 \centering{\includegraphics[trim=0 80bp 0 80bp,clip,scale=.5]{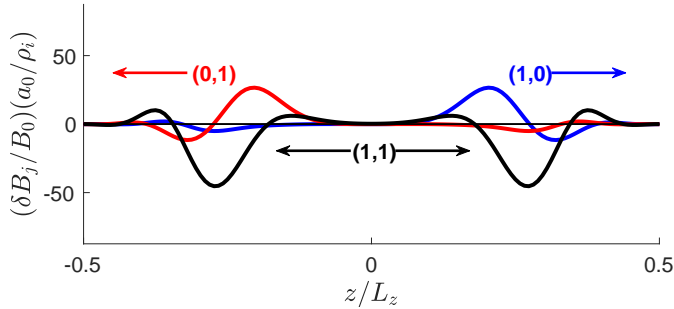}}
 \caption{Snapshots in time of $\delta B_\perp$ vs. $z$ of primary modes (1,0) and (0,1) overlapping the secondary mode (1,1) in the \T{LS} case.  All times are normalized to the \Alfven collision time, $T_c^{(l)}$.
 \label{fig:disp}}
 \end{figure}

A brute-force determination of whether any nonlinearly generated mode satisfies the linear \Alfven wave dispersion relation requires a decomposition of the fluctuation into plane-wave modes to enable a comparison between the parallel wavenumber $k_\parallel$ of each constituent plane-wave mode and its linear frequency $\omega$.  Such a task is complicated for the case of collisions between localized \Alfven wavepackets, which necessarily contain a broad spectrum of parallel wavenumbers to accomplish localization in $z$.  But the non-dispersive nature of \Alfven waves makes an alternative approach possible: if the nonlinearly generated modes propagate along the equilibrium field direction together with the original \Alfven wavepackets at the \Alfven speed, then collectively they describe a localized wavepacket propagating non-dispersively.  In \figref{fig:disp}, we overplot the perpendicular magnetic field perturbation $\delta B_\perp$ of the secondary (1,1) Fourier mode with that of the primary (0,1) and (1,0) Fourier modes at times $t/T^{(l)}_c= 1,2,3$, showing that the nonlinearly generated (1,1) mode does indeed propagate up or down along $z$ with the primary modes at the \Alfven speed.  Furthermore, as predicted from the analytical solution for \Alfven wave collisions \citep{Howes:2013a, Howes:2013b}, the (1,1) mode is phase shifted by $\pi/2$ relative to the primary mode from which it gained energy.  For example, in \figref{fig:disp}(a), the downward (0,1) mode (red) passes through zero at the same position in $z$ at which the downward propagating secondary (1,1) mode (black) reaches a peak.  The crucial point of \figref{fig:disp} is that, in the localized \Alfven wavepacket collision, the nonlinearly generated, secondary (1,1) Fourier mode satisfies the linear \Alfven wave dispersion relation, propagating along the equilibrium magnetic field non-dispersively.

It is worthwhile noting that the gyrokinetic simulations performed here indeed captures the physics of the finite-ion-Larmor-radius corrections that cause the \Alfven wave solution to become dispersive at $k_\perp \rho_i \rightarrow 1$, transitioning to the dispersive kinetic \Alfven wave.  Therefore, there is a very slight spreading of the wavepackets after nonlinear interactions have transferred energy into modes with $k_\perp \rho_i \gtrsim 1$.  This behavior is noticeable in Figure~7 of our companion paper \cite{Verniero:2017a} and is discussed in more detail in Section 3.4 of that paper.

In summary, the results presented in \figref{fig:eig} and \figref{fig:els} show that, in the more realistic strong, localized \Alfven wavepacket collision case,  the secondary (1,1) mode satisfies the linear \Alfven wave eigenfunction condition.  The results presented in \figref{fig:disp} show that this mode also satisfies the linear \Alfven wave dispersion relation.  Therefore, we conclude that this secondary (1,1) Fourier mode, which plays a key role in the nonlinear transfer of energy to smaller perpendicular scales, is simply an \Alfven wave.  Note that one may interpret this (1,1) mode of the \Alfven wave as a shear that propagates along the magnetic field at the \Alfven speed \citep{Howes:2017b}. This finding leads to a simplification of the picture of the nonlinear cascade of energy in plasma turbulence relative to the idealized (but analytically soluble) periodic case.  In the periodic case, the nonlinear energy transfer to smaller scales was mediated by an inherently nonlinear (1,1) Fourier mode.  In the more realistic localized case, the energy mode that mediates the energy transfer is simply an \Alfven wave itself, both gaining energy from the nonlinear interaction and mediating further energy transfer to smaller scales.

\subsection{Strong vs. Weak Turbulence}
\label{sec:sw}

Although the primary aim of this study is to understand how the physics of \Alfven wave collisions changes in the more realistic case of localized \Alfven wavepacket collisions, it is also worthwhile to explore the differences between the weak and strong cases in both the periodic and localized cases.

In \figref{fig:energy3col} and \figref{fig:fullenergy}, comparing the (d) weakly and (b) strongly nonlinear periodic cases, the most obvious difference is that the energy of the primary \Alfven waves is significantly diminished in the strongly nonlinear case, whereas in the weakly nonlinear case, the loss of energy by the primary \Alfven waves is negligible, even over the long time scale shown in \figref{fig:fullenergy}(d), as expected.  What is not necessarily expected is that the evolution between the strongly and weakly nonlinear periodic cases is qualitatively similar, with the secondary (1,1) mode and the tertiary (1,2) and (2,1) \Alfven waves as the dominant recipients of the energy nonlinearly transferred from the primary \Alfven waves.  The physics governing the nonlinear cascade of energy to smaller scales appears to be similar in the weakly and strongly nonlinear limits, suggesting that physical intuition from the weakly nonlinear limit provides a useful framework for the interpretation of the strongly nonlinear dynamics.  Such an approach, in fact, underlies the recent discovery that strong \Alfven wave collisions naturally develop current sheets \citep{Howes:2016b}.  A final qualitative feature of the long term evolution in the \T{PS} case, shown in \figref{fig:fullenergy}(b), is that the primary \Alfven waves lose energy up to $t/T^{(p)}_c \sim 5$, and then their amplitudes begin to rise again.  This curious behavior arises from the dispersive nature of kinetic \Alfven waves in the limit $k_\perp \rho_i \rightarrow 1$.  The nonlinearly generated tertiary \Alfven waves in the gyrokinetic system have slight dispersive increase in their frequency due finite Larmor radius averaging, and over time begin to shift out of phase with the primary modes, eventually transferring some of their energy back to the primary waves \citep{Nielson:2012}.

Comparing the (c) weakly and (a) strongly nonlinear localized cases in \figref{fig:energy3col} and \figref{fig:fullenergy},  we observe the same qualitative similarity between the weakly and strongly nonlinear dynamics, with a more significant fraction of energy lost by the primary \Alfven wavepackets in the strongly nonlinear case, again as expected.  In contrast to the periodic cases, in both weakly and strongly nonlinear localized cases, all nonlinearly generated modes gain energy secularly over time.  Because all of these smaller perpendicular scale modes are gaining energy, there is a substantially greater loss of energy from the primary \Alfven wavepackets in the \T{LW} case relative to the loss from the primary \Alfven waves in the \T{PW} case, clearly shown by comparing \figref{fig:fullenergy}(c) and (d).  The strongly nonlinear \T{LS} and \T{PS} cases in \figref{fig:fullenergy}(a) and (b) show a similar relation, where the energy loss from the localized case is much more significant than in the periodic case.  Therefore, it appears that localized \Alfven wavepacket collisions are much more effective in mediating the nonlinear cascade to smaller perpendicular scales.  This is a key result because the localized, strongly nonlinear \T{LS} case,  the primary focus of this paper, is the most physically relevant case for application to particular space and astrophysical environments, such as the solar wind and solar corona.
\subsection{Current Sheet Development}
\label{sec:cur}	

\begin{figure}
\hspace{0.1in} (a) $t/T_c^{(l)}$=0 at -1/4$L_z$ \hspace{1.45 in}  (b) $t/T_c^{(l)}$=0 at 1/4$L_z$
\vfill
	{\includegraphics[scale=.45]{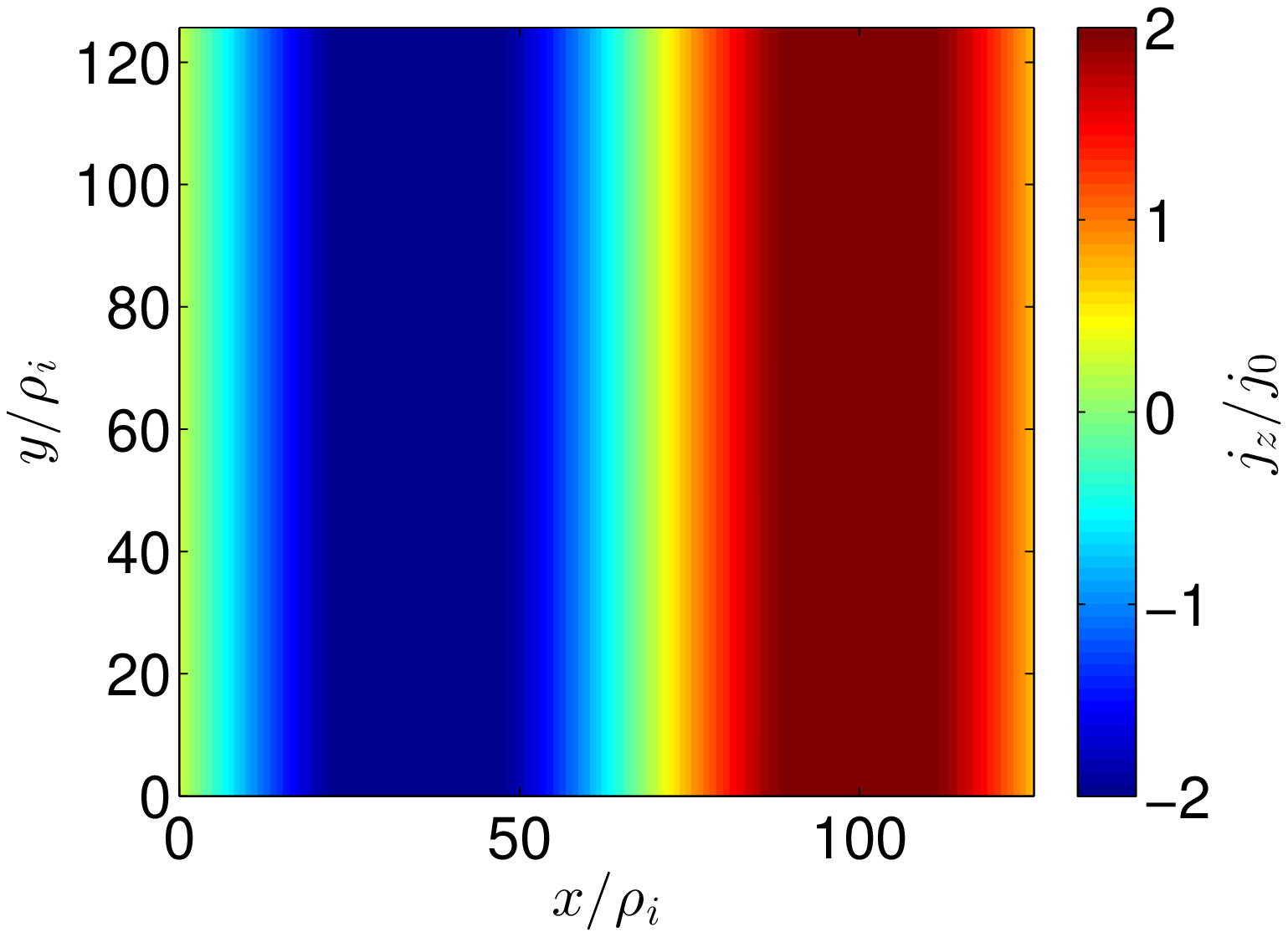}}\hfill
	{\includegraphics[scale=.45]{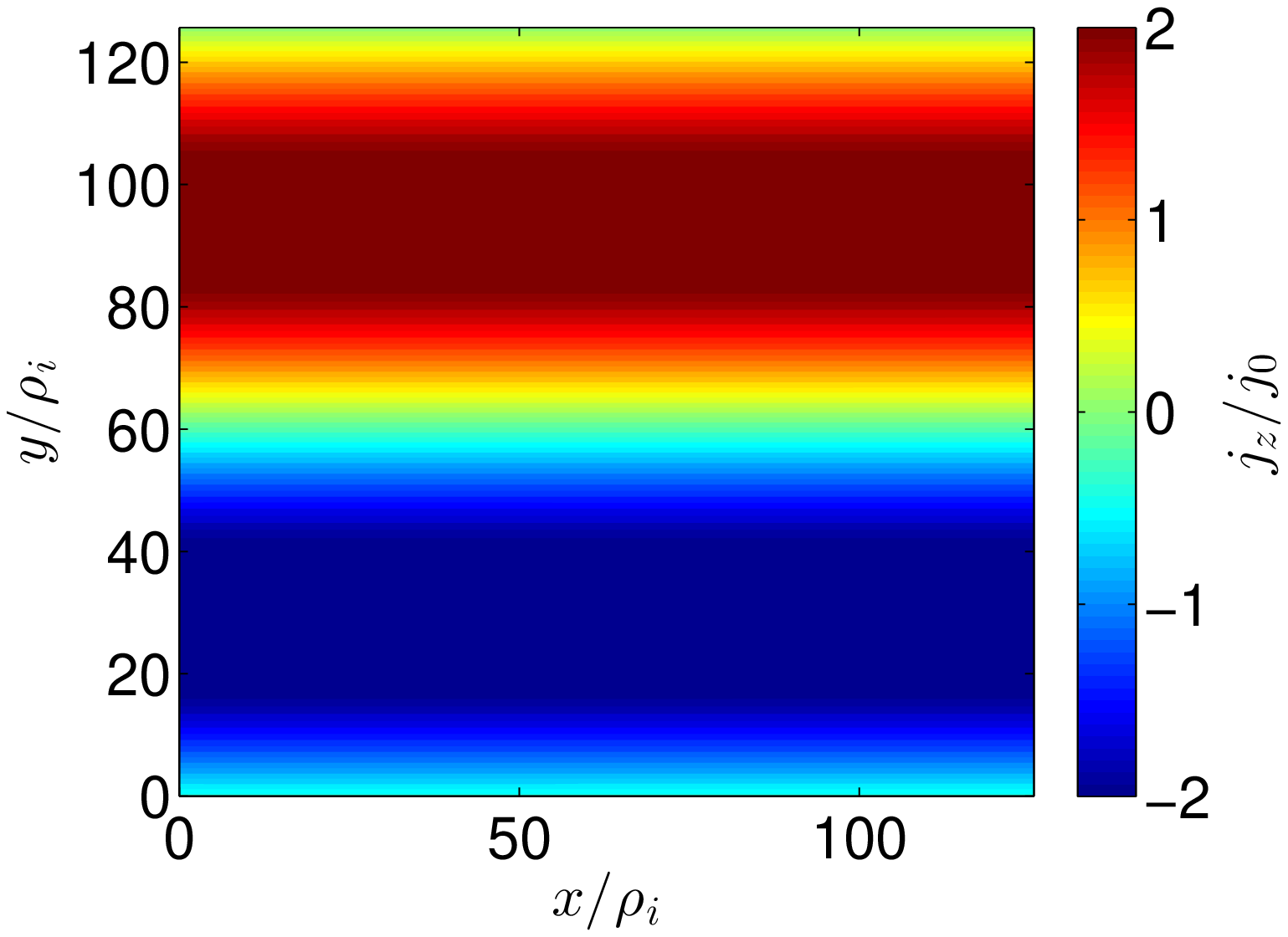}}
\vfill
\hspace{0.1in} (c) $t/T_c^{(l)}$=1 at 1/4$L_z$ \hspace{1.5 in}  (d) $t/T_c^{(l)}$=1 at -1/4$L_z$
\vfill
	{\includegraphics[scale=.45]{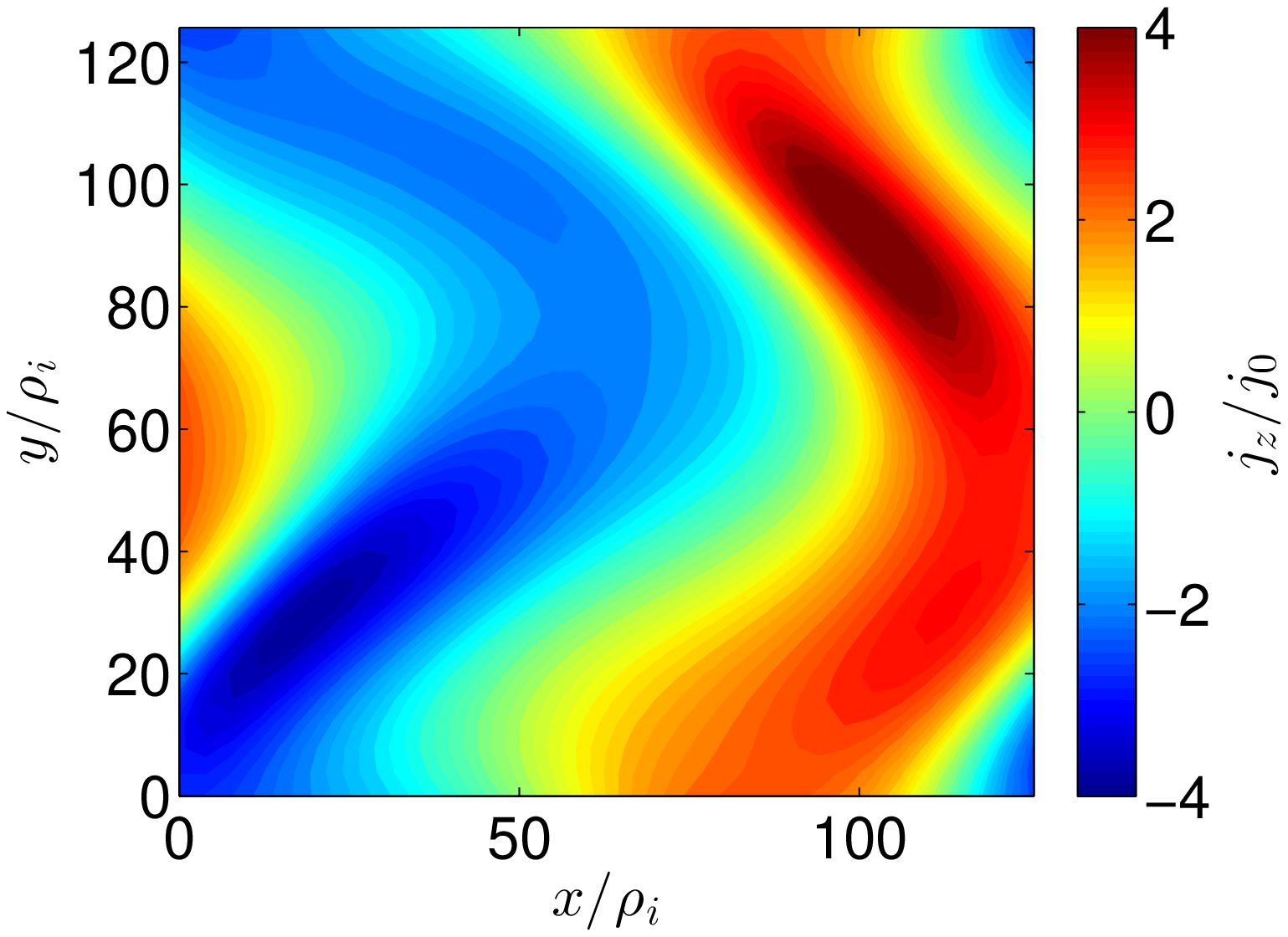}}\hfill
	{\includegraphics[scale=.45]{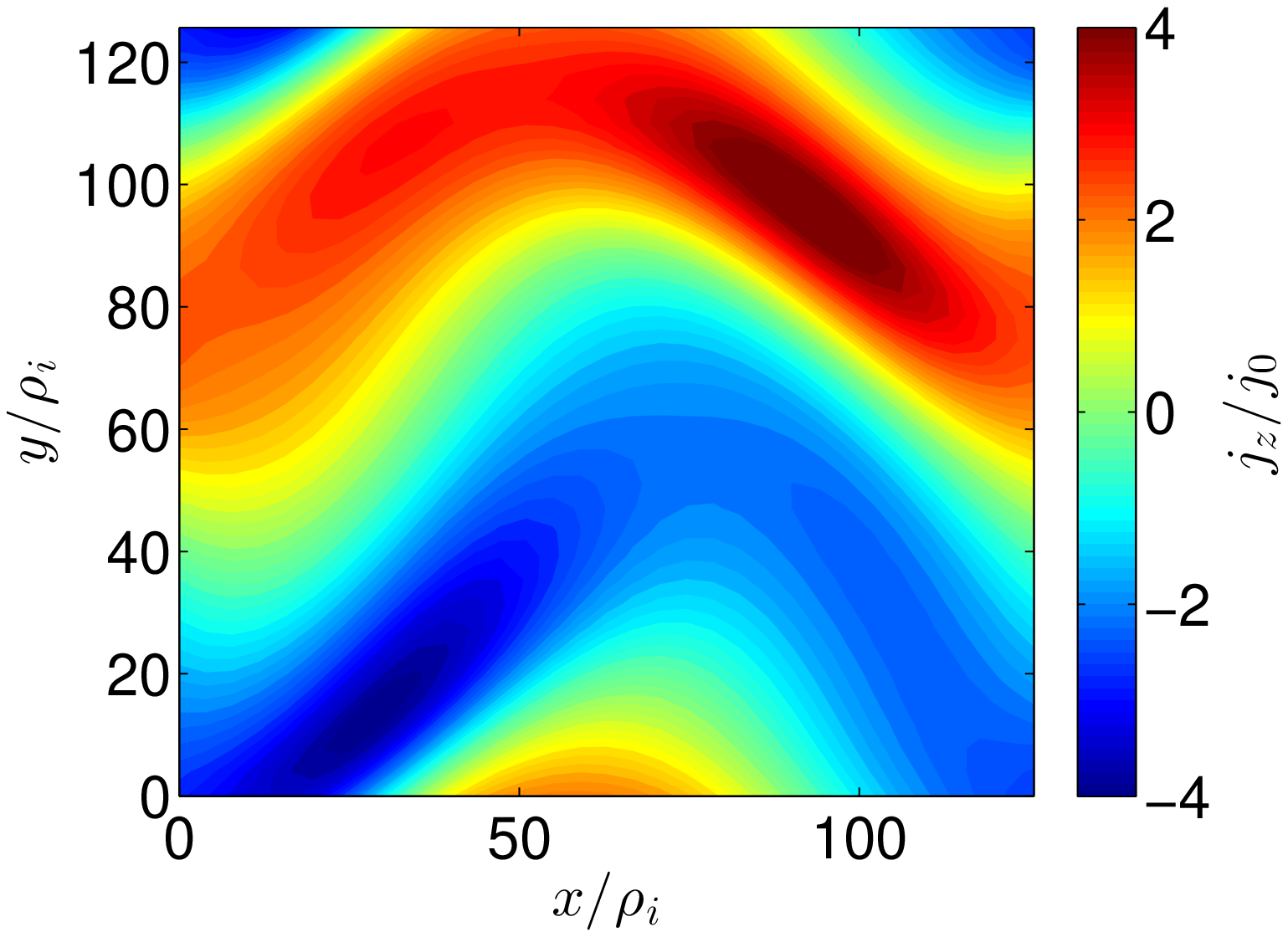}}
\vfill
\hspace{0.1in} (e) $t/T_c^{(l)}$=2 at -1/4$L_z$ \hspace{1.5in}  (f) $t/T_c^{(l)}$=2 at 1/4$L_z$
\vfill
        {\includegraphics[scale=.45]{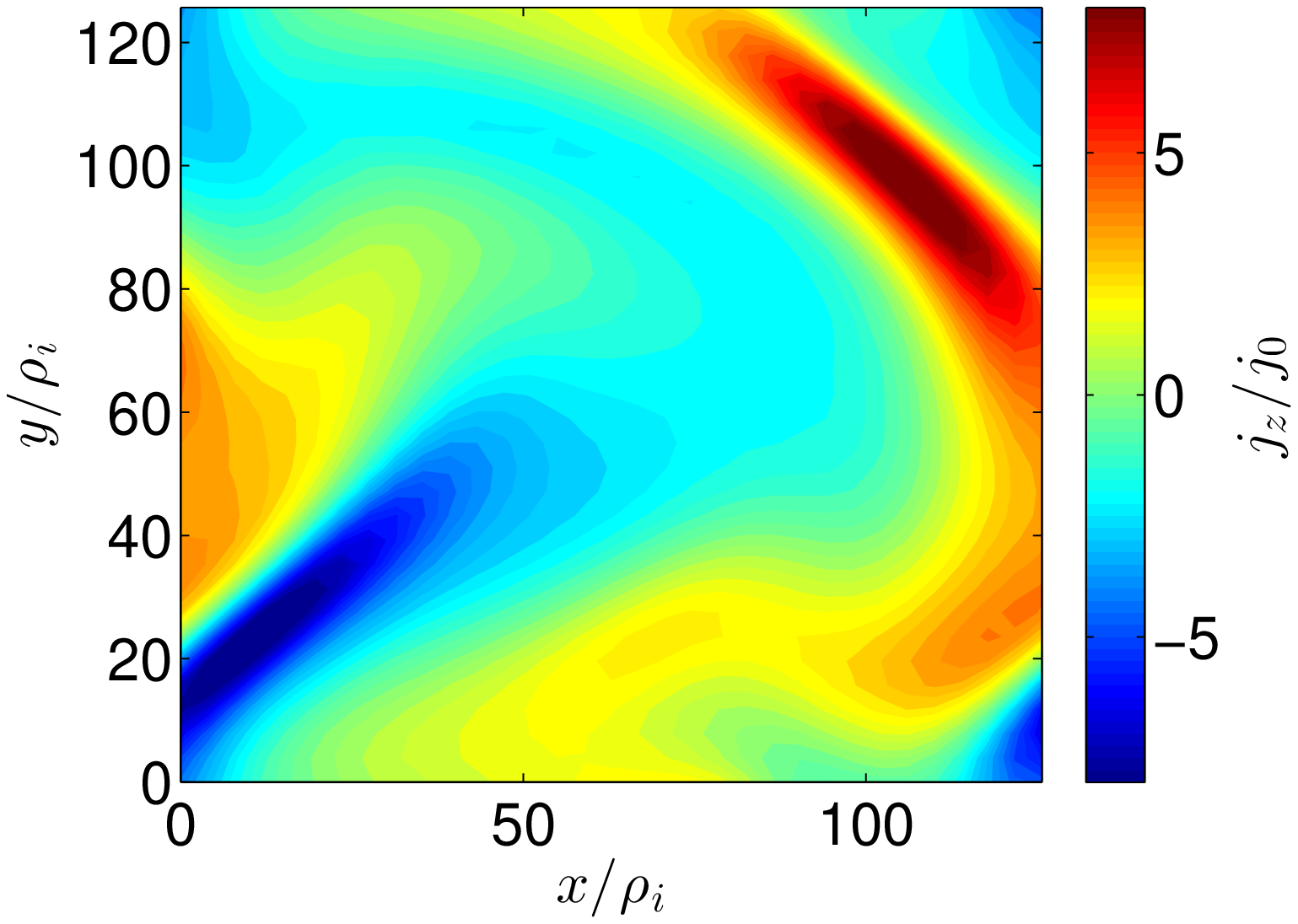}}\hfill	
        {\includegraphics[scale=.45]{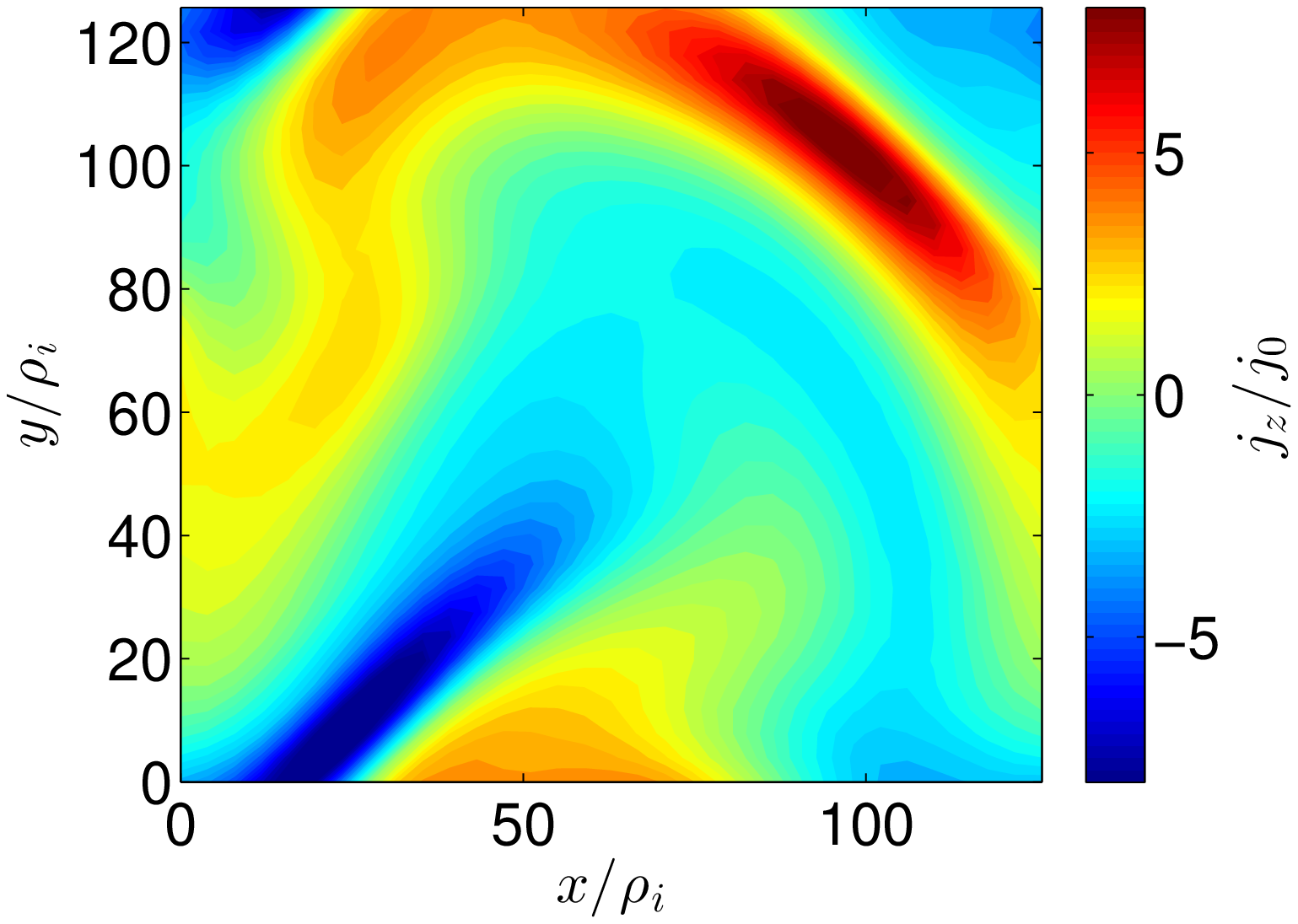}}        
\caption{Current sheet formation before and after each collision of case \T{LS}. 
\label{fig:current}}	
\end{figure}

The final aim of this paper to determine whether current sheets naturally develop in the case of a strongly nonlinear collision between two symmetric \Alfven wavepackets, where neither initially has a substantial $k_\parallel =0$ component.  \figref{fig:current} shows plots of the normalized parallel current density $j_z/j_0 = (j_z/n_{0i} q_i v_{ti}) (a_0/\rho_0)$ in the $(x, y)$ plane perpendicular to the equilibrium magnetic field.

The left column of \figref{fig:current} follows the evolution of the upward propagating $z^-$ \Alfven wavepacket, while the right column shows the downward propagating $z^+$ \Alfven wavepacket. Note that the waves collide at the midpoint of the simulation box $z=0$ and periodically at the end points $z= \pm L_z/2$.  We plot the perpendicular cross section of the parallel current density $j_z$ of each wavepacket at $z=\pm L_z/4$ when the wavepackets are not overlapping at $t=0$ in (a) and (b), after the first collision at $t/T^{(l)}_c=1$ in (c) and (d), and after the second collision at $t/T^{(l)}_c=2$ in (e) and (f).  In (c) and (d), we see that the nonlinear distortion of the original current pattern persists after the first collision, leading to a narrowing and intensification of the current sheet.  After the second collision in (e) and (f), the current density has further thinned and intensified into a sheet-like morphology.  Note that the amplitude of the color scale increases with later snapshots, making it clear that the current sheets are becoming increasingly intense and narrow over time. Therefore, the result first shown in \cite{Verniero:2017a}, that strong localized \Alfven wavepacket collisions naturally lead to the development of current sheets, is not dependent on the nonzero $k_\parallel =0$ component of one of the colliding \Alfven wavepackets in that study. We may therefore conclude that the development of current sheets in strong, localized \Alfven wavepacket collisions is a robust result that is not dependent on any particular forms of the initial wavepackets, further extending the impact of the initial discovery that strong \Alfven wave collisions self-consistently generate current sheets \citep{Howes:2016b}, providing a first-principles explanation for the ubiquitous observations of current sheets in turbulent space and astrophysical plasmas.

\section{Conclusion}
\label{sec:con}

The results presented in this paper settles the issue of the nature of the nonlinearly generated secondary mode -- the mode that mediates the nonlinear transfer of energy in \Alfven wave collisions -- in a more realistic setting than the idealized periodic case that was used in previous work to enable an analytical solution to be computed.  Addressing the first question in the introduction, we conclude that these secondary modes are indeed \Alfven modes in the case of localized \Alfven wavepacket collisions. This fact was confirmed by showing (i) the eigenfunction condition, that there is the correct relationship between the $\mathbf{E}$ and $\mathbf{B}$ fields described by Equation \ref{eq:eig} and shown in \figref{fig:eig} and \figref{fig:els} and (ii) the correct frequency condition, that the (1,1) mode travels at the \Alfven speed in accordance with the rest of the energy modes as shown in \figref{fig:disp}.

 Observing \figref{fig:energy3col} and \figref{fig:fullenergy}, we found that in the periodic cases, only the tertiary (1,2) and (2,1) modes experience a secular gain of energy after successive collisions, while in the localized cases, the secondary (1,1) mode gains energy in addition to the tertiary modes. This means that in the case of localized wave collisions in both the strongly and weakly nonlinear limit, energy transfer to smaller perpendicular scales is more efficient than in the periodic case. We also saw, by comparison between the weakly and strongly nonlinear cases, that the primary modes in the strongly nonlinear limit lose significantly more energy than the weakly nonlinear cases.  This saturation is the most discernible quantitative difference between strong and weak turbulence, while most other key features remain qualitatively similar such as overall evolution of the energy of different perpendicular Fourier modes in time. We conclude that the strong, localized \T{LS} case is the most effective way to transfer energy to smaller perpendicular scales.  This particular case of localized, strong turbulence, the focal point of this paper, is the most applicable case to space and astrophysical plasmas. Hence, this case is crucial for understanding the various turbulent energy cascades within our universe such as black hole accretion disks, the solar wind, and planetary magnetospheres. 

From \figref{fig:current}, we have also demonstrated that for the strong, localized \T{LS} case, self-consistent current sheets are generated after successive collisions and persist in between collisions, consistent with previous findings on strong turbulence simulations for both the periodic \citep{Howes:2016b} and initially asymmetric localized cases \citep{Verniero:2017a}. This particular finding shows that \Alfven wavepacket collisions in the strongly nonlinear limit are a robust mechanism for current sheet development, regardless of initial waveform.  In turbulent space and astrophysical plasmas, current sheets are observed ubiquitously and have been proposed to play a key role in the conversion of turbulent energy into plasma heat.  The quest to understand how a plasma becomes heated is currently an active topic of research in the plasma physics community.  For example, the \emph{Parker Solar Probe}, due to be launched in July 2018, will investigate how the solar corona becomes heated to unprecedented temperatures, a topic that has been debated for decades.  The result presented in this paper -- that localized, strongly nonlinear \Alfven wave collisions naturally produce current sheets -- means that the observations of current sheets in many space and astrophysical plasma systems can be explained from first principles. 

We conclude that in the most physically applicable case of localized, strongly nonlinear interactions, the fundamental properties of plasma turbulence still persist: energy cascades nonlinearly to smaller perpendicular scales and intermittent current sheets are self-consistently generated, answering the second question posed in the introduction.  In \citet{Verniero:2017a}, we analyzed the case of localized, strongly nonlinear \Alfven wavepacket collisions with asymmetric initial waveforms.  The symmetric conditions presented in this paper demonstrate clearly that the effect of a nonzero $k_\parallel$ component does not alter the main characteristics of the \Alfven wave collisions that govern plasma turbulence. 
 
Our findings of the \Alfvenic nature of the key (1,1) mode in the localized, strongly nonlinear case is a satisfying simplification of the picture of the nonlinear energy transfer to small scales in plasma turbulence.  It is important to emphasize the fact that an \Alfven wave collision is the fundamental unit of interaction in plasma turbulence \citep{Kraichnan:1965,Howes:2013a}, and a turbulent plasma would contain many such nonlinear interactions among upward and downward propagating \Alfven wavepackets.  Such an \emph{ab initio} approach to this subject allows for a clearer picture to be painted and consequently enables deeper insight about the dynamics. The results presented in this paper highlight the central role played by \Alfven waves in the nonlinear cascade of energy.  The generation of the secondary mode mediates the transfer of energy from the primary to tertiary modes. The secondary mode is essentially a shear in the magnetic field that propagates along the magnetic field as an \Alfven wave, shearing the perpendicular waveform of counterpropagating \Alfven wavepackets and thereby nonlinearly transferring their energy to smaller perpendicular scales \citep{Howes:2017b}.  In contrast to the idealized periodic case, this secondary (1,1) mode gains energy secularly along with all of the other nonlinearly generated modes. The striking difference between the periodic case with two initially overlapping plane \Alfven waves and the localized \Alfven wavepacket case raises the question of whether the non-\Alfvenic ``beat" modes that arise in the periodic case will alter the statistics of the turbulence.  For decaying turbulence simulations, in which the initialized \Alfvenic fluctuations are already overlapping as in our periodic case, this is an issue that merits further investigation.
 A follow up study could investigate the role of the \Alfvenic propagating shear, discussed in this paper, on magnetic field line wander, enabling a more atomistic description of the tangling of magnetic field lines within the framework of \Alfven wave collisions. Our analysis of the more realistic case of localized \Alfven wave collisions brings us closer to understanding the fundamental characteristics of plasma turbulence from first principles.


This material is based upon work supported by the National Science Foundation Graduate Research Fellowship Program under Grant No. 1048957, NSF PHY-10033446, NSF CAREER AGS-1054061, and DOE
DE-SC0014599.  This research used resources of the Oak Ridge
Leadership Computing Facility, which is a DOE Office of Science User
Facility supported under Contract DE-AC05-00OR22725. This work used the Extreme Science and
Engineering Discovery Environment (XSEDE), which is supported by
National Science Foundation grant number ACI-1053575, through NSF
XSEDE Award PHY090084. 




\begin{thebibliography}{41}
\expandafter\ifx\csname natexlab\endcsname\relax\def\natexlab#1{#1}\fi

\bibitem[{Abel} {\em et~al.\/}(2008){Abel}, {Barnes}, {Cowley}, {Dorland} \&
  {Schekochihin}]{Abel:2008}
{\sc {Abel}, I.~G., {Barnes}, M., {Cowley}, S.~C., {Dorland}, W. \&
  {Schekochihin}, A.~A.} 2008 {Linearized model Fokker-Planck collision
  operators for gyrokinetic simulations. I. Theory}. {\em Phys.~Plasmas\/} {\bf                                                                                                                              
  15}~(12), 122509.

\bibitem[{Barnes} {\em et~al.\/}(2009){Barnes}, {Abel}, {Dorland}, {Ernst},
  {Hammett}, {Ricci}, {Rogers}, {Schekochihin} \& {Tatsuno}]{Barnes:2009}
{\sc {Barnes}, M., {Abel}, I.~G., {Dorland}, W., {Ernst}, D.~R., {Hammett},
  G.~W., {Ricci}, P., {Rogers}, B.~N., {Schekochihin}, A.~A. \& {Tatsuno}, T.}
  2009 {Linearized model Fokker-Planck collision operators for gyrokinetic
  simulations. II. Numerical implementation and tests}. {\em Phys.~Plasmas\/}
  {\bf 16}~(7), 072107.

\bibitem[{Borovsky} \& {Denton}(2011)]{Borovsky:2011}
{\sc {Borovsky}, J.~E. \& {Denton}, M.~H.} 2011 {No Evidence for Heating of the
  Solar Wind at Strong Current Sheets}. {\em Astrophys.~J.~Lett.\/} {\bf 739},
  L61.

\bibitem[{Drake} {\em et~al.\/}(2016){Drake}, {Howes}, {Rhudy}, {Terry},
  {Carter}, {Kletzing}, {Schroeder} \& {Skiff}]{Drake:2016}
{\sc {Drake}, D.~J., {Howes}, G.~G., {Rhudy}, J.~D., {Terry}, S.~K., {Carter},
  T.~A., {Kletzing}, C.~A., {Schroeder}, J.~W.~R. \& {Skiff}, F.} 2016
  {Measurements of the nonlinear beat wave produced by the interaction of
  counterpropagating Alfven waves}. {\em Phys.~Plasmas\/} {\bf 23}~(2), 022305.

\bibitem[{Drake} {\em et~al.\/}(2013){Drake}, {Schroeder}, {Howes}, {Kletzing},
  {Skiff}, {Carter} \& {Auerbach}]{Drake:2013}
{\sc {Drake}, D.~J., {Schroeder}, J.~W.~R., {Howes}, G.~G., {Kletzing}, C.~A.,
  {Skiff}, F., {Carter}, T.~A. \& {Auerbach}, D.~W.} 2013 {Alfv{\'e}n wave
  collisions, the fundamental building block of plasma turbulence. IV.
  Laboratory experiment}. {\em Physics of Plasmas\/} {\bf 20}~(7), 072901.

\bibitem[{Frieman} \& {Chen}(1982)]{Frieman:1982}
{\sc {Frieman}, E.~A. \& {Chen}, L.} 1982 {Nonlinear gyrokinetic equations for
  low-frequency electromagnetic waves in general plasma equilibria}. {\em                                                                                                                                    
  Phys.~Fluids\/} {\bf 25}, 502--508.

\bibitem[{Galtier} {\em et~al.\/}(2000){Galtier}, {Nazarenko}, {Newell} \&
  {Pouquet}]{Galtier:2000}
{\sc {Galtier}, S., {Nazarenko}, S.~V., {Newell}, A.~C. \& {Pouquet}, A.} 2000
  {A weak turbulence theory for incompressible magnetohydrodynamics}. {\em                                                                                                                                   
  J.~Plasma Phys.\/} {\bf 63}, 447--488.

\bibitem[Goldreich \& Sridhar(1995)]{Goldreich:1995}
{\sc Goldreich, P. \& Sridhar, S.} 1995 {Toward a Theery of Interstellar
  Turbulence II. Strong Alfv\'enic Turbulence}. {\em Astrophys.~J.\/} {\bf                                                                                                                                   
  438}, 763--775.

\bibitem[{Howes}(2016)]{Howes:2016b}
{\sc {Howes}, G.~G.} 2016 {The Dynamical Generation of Current Sheets in
  Astrophysical Plasma Turbulence}. {\em Astrophys.~J.~Lett.\/} {\bf 827}, L28.

\bibitem[{Howes}(2017)]{Howes:2017c}
{\sc {Howes}, G.~G.} 2017 {A Prospectus on Kinetic Heliophysics}. {\em                                                                                                                                       
  Phys.~Plasmas\/} In press.

\bibitem[{Howes} \& {Bourouaine}(2017)]{Howes:2017b}
{\sc {Howes}, G.~G. \& {Bourouaine}, S.} 2017 {The Development of Magnetic
  Field Line Wander by Plasma Turbulence}. {\em J.~Plasma Phys.\/} Submitted.
  
  \bibitem[{Howes} {\em et~al.\/}(2006){Howes}, {Cowley}, {Dorland}, {Hammett},
  {Quataert} \& {Schekochihin}]{Howes:2006}
{\sc {Howes}, G.~G., {Cowley}, S.~C., {Dorland}, W., {Hammett}, G.~W.,
  {Quataert}, E. \& {Schekochihin}, A.~A.} 2006 {Astrophysical Gyrokinetics:
  Basic Equations and Linear Theory}. {\em Astrophys.~J.\/} {\bf 651},
  590--614.

\bibitem[{Howes} {\em et~al.\/}(2012){Howes}, {Drake}, {Nielson}, {Carter},
  {Kletzing} \& {Skiff}]{Howes:2012b}
{\sc {Howes}, G.~G., {Drake}, D.~J., {Nielson}, K.~D., {Carter}, T.~A.,
  {Kletzing}, C.~A. \& {Skiff}, F.} 2012 {Toward Astrophysical Turbulence in
  the Laboratory}. {\em Phys.~Rev.~Lett.\/} {\bf 109}~(25), 255001.

\bibitem[{Howes} {\em et~al.\/}(2017{\natexlab{{\em a\/}}}){Howes}, {Klein} \&
  {Li}]{Howes:2017a}
{\sc {Howes}, G.~G., {Klein}, K.~G. \& {Li}, T.~C.} 2017{\natexlab{{\em a\/}}}
  {Diagnosing Collisionless Energy Transfer Using Wave-Particle Correlations:
  Vlasov-Poisson Plasmas}. {\em J.~Plasma Phys.\/} {\bf 83}~(1), 705830102.

\bibitem[{Howes} {\em et~al.\/}(2017{\natexlab{{\em b\/}}}){Howes}, {McCubbin}
  \& {Klein}]{Howes:2017d}
{\sc {Howes}, G.~G., {McCubbin}, A.~J. \& {Klein}, K.~G.} 2017{\natexlab{{\em                                                                                                                                
  b\/}}} {Spatial Localization of Particle Enegization in Current Sheets
  Produced by \Alfven Wave Collisions}. {\em J.~Plasma Phys.\/} Submitted.

\bibitem[{Howes} \& {Nielson}(2013)]{Howes:2013a}
{\sc {Howes}, G.~G. \& {Nielson}, K.~D.} 2013 {Alfv{\'e}n wave collisions, the
  fundamental building block of plasma turbulence. I. Asymptotic solution}.
  {\em Phys.~Plasmas\/} {\bf 20}~(7), 072302.

\bibitem[{Howes} {\em et~al.\/}(2013){Howes}, {Nielson}, {Drake}, {Schroeder},
  {Skiff}, {Kletzing} \& {Carter}]{Howes:2013b}
{\sc {Howes}, G.~G., {Nielson}, K.~D., {Drake}, D.~J., {Schroeder}, J.~W.~R.,
  {Skiff}, F., {Kletzing}, C.~A. \& {Carter}, T.~A.} 2013 {Alfv{\'e}n wave
  collisions, the fundamental building block of plasma turbulence. III. Theory
  for experimental design}. {\em Physics of Plasmas\/} {\bf 20}~(7), 072304.

\bibitem[Iroshnikov(1963)]{Iroshnikov:1963}
{\sc Iroshnikov, R.~S.} 1963 The turbulence of a conducting fluid in a strong
  magnetic field. {\em Astron. Zh.\/} {\bf 40}, 742, {English} Translation:
  Sov. Astron., 7 566 (1964).

\bibitem[{Karimabadi} {\em et~al.\/}(2013){Karimabadi}, {Roytershteyn}, {Wan},
  {Matthaeus}, {Daughton}, {Wu}, {Shay}, {Loring}, {Borovsky}, {Leonardis},
  {Chapman} \& {Nakamura}]{Karimabadi:2013}
{\sc {Karimabadi}, H., {Roytershteyn}, V., {Wan}, M., {Matthaeus}, W.~H.,
  {Daughton}, W., {Wu}, P., {Shay}, M., {Loring}, B., {Borovsky}, J.,
  {Leonardis}, E., {Chapman}, S.~C. \& {Nakamura}, T.~K.~M.} 2013 {Coherent
  structures, intermittent turbulence, and dissipation in high-temperature
  plasmas}. {\em Phys.~Plasmas\/} {\bf 20}~(1), 012303.

\bibitem[{Klein} \& {TenBarge}(2017)]{Klein:2017b}
{\sc {Klein}, K.~G.and~{Howes}, G.~G. \& {TenBarge}, J.~M.} 2017 {Diagnosing
  collisionless energy transfer using field-particle correlations: gyrokinetic
  turbulence}. {\em J.~Plasma Phys.\/} Submitted.

\bibitem[{Klein} \& {Howes}(2016)]{Klein:2016a}
{\sc {Klein}, K.~G. \& {Howes}, G.~G.} 2016 {Measuring Collisionless Damping in
  Heliospheric Plasmas using Field-Particle Correlations}. {\em                                                                                                                                              
  Astrophys.~J.~Lett.\/} {\bf 826}, L30.

\bibitem[Kraichnan(1965)]{Kraichnan:1965}
{\sc Kraichnan, R.~H.} 1965 Inertial range spectrum of hyromagnetic turbulence.
  {\em Phys.~Fluids\/} {\bf 8}, 1385--1387.

\bibitem[{Matthaeus} \& {Montgomery}(1980)]{Matthaeus:1980}
{\sc {Matthaeus}, W.~H. \& {Montgomery}, D.} 1980 {Selective decay hypothesis
  at high mechanical and magnetic Reynolds numbers}. {\em Annals of the New                                                                                                                                  
  York Academy of Sciences\/} {\bf 357}, 203--222.

\bibitem[{Meneguzzi} {\em et~al.\/}(1981){Meneguzzi}, {Frisch} \&
  {Pouquet}]{Meneguzzi:1981}
{\sc {Meneguzzi}, M., {Frisch}, U. \& {Pouquet}, A.} 1981 {Helical and
  nonhelical turbulent dynamos}. {\em Phys.~Rev.~Lett.\/} {\bf 47}, 1060--1064.

\bibitem[{Montgomery} \& {Matthaeus}(1995)]{Montgomery:1995}
{\sc {Montgomery}, D. \& {Matthaeus}, W.~H.} 1995 {Anisotropic Modal Energy
  Transfer in Interstellar Turbulence}. {\em Astrophys.~J.\/} {\bf 447}, 706.

\bibitem[{Ng} \& {Bhattacharjee}(1996)]{Ng:1996}
{\sc {Ng}, C.~S. \& {Bhattacharjee}, A.} 1996 {Interaction of Shear-Alfven Wave
  Packets: Implication for Weak Magnetohydrodynamic Turbulence in Astrophysical
  Plasmas}. {\em Astrophys.~J.\/} {\bf 465}, 845.

\bibitem[{Nielson}(2012)]{Nielson:2012}
{\sc {Nielson}, K.~D.} 2012 {Analysis and gyrokinetic simulation of MHD Alfven
  wave interactions}. PhD thesis, The University of Iowa.

\bibitem[{Nielson} {\em et~al.\/}(2013){Nielson}, {Howes} \&
  {Dorland}]{Nielson:2013a}
{\sc {Nielson}, K.~D., {Howes}, G.~G. \& {Dorland}, W.} 2013 {Alfv{\'e}n wave
  collisions, the fundamental building block of plasma turbulence. II.
  Numerical solution}. {\em Physics of Plasmas\/} {\bf 20}~(7), 072303.

\bibitem[{Numata} {\em et~al.\/}(2010){Numata}, {Howes}, {Tatsuno}, {Barnes} \&
  {Dorland}]{Numata:2010}
{\sc {Numata}, R., {Howes}, G.~G., {Tatsuno}, T., {Barnes}, M. \& {Dorland},
  W.} 2010 {AstroGK: Astrophysical gyrokinetics code}. {\em J.~Comp.~Phys.\/}
  {\bf 229}, 9347.
  
  \bibitem[{Osman} {\em et~al.\/}(2014){Osman}, {Matthaeus}, {Gosling}, {Greco},
  {Servidio}, {Hnat}, {Chapman} \& {Phan}]{Osman:2014b}
{\sc {Osman}, K.~T., {Matthaeus}, W.~H., {Gosling}, J.~T., {Greco}, A.,
  {Servidio}, S., {Hnat}, B., {Chapman}, S.~C. \& {Phan}, T.~D.} 2014 {Magnetic
  Reconnection and Intermittent Turbulence in the Solar Wind}. {\em                                                                                                                                          
  Phys.~Rev.~Lett.\/} {\bf 112}~(21), 215002.

\bibitem[{Osman} {\em et~al.\/}(2011){Osman}, {Matthaeus}, {Greco} \&
  {Servidio}]{Osman:2011}
{\sc {Osman}, K.~T., {Matthaeus}, W.~H., {Greco}, A. \& {Servidio}, S.} 2011
  {Evidence for Inhomogeneous Heating in the Solar Wind}. {\em                                                                                                                                               
  Astrophys.~J.~Lett.\/} {\bf 727}, L11.

\bibitem[{Osman} {\em et~al.\/}(2012){Osman}, {Matthaeus}, {Wan} \&
  {Rappazzo}]{Osman:2012a}
{\sc {Osman}, K.~T., {Matthaeus}, W.~H., {Wan}, M. \& {Rappazzo}, A.~F.} 2012
  {Intermittency and Local Heating in the Solar Wind}. {\em Phys.~Rev.~Lett.\/}
  {\bf 108}~(26), 261102.

\bibitem[{Perri} {\em et~al.\/}(2012){Perri}, {Goldstein}, {Dorelli} \&
  {Sahraoui}]{Perri:2012a}
{\sc {Perri}, S., {Goldstein}, M.~L., {Dorelli}, J.~C. \& {Sahraoui}, F.} 2012
  {Detection of Small-Scale Structures in the Dissipation Regime of Solar-Wind
  Turbulence}. {\em Phys.~Rev.~Lett.\/} {\bf 109}~(19), 191101.

\bibitem[{Sridhar} \& {Goldreich}(1994)]{Sridhar:1994}
{\sc {Sridhar}, S. \& {Goldreich}, P.} 1994 {Toward a theory of interstellar
  turbulence. 1: Weak Alfvenic turbulence}. {\em Astrophys.~J.\/} {\bf 432},
  612--621.

\bibitem[{TenBarge} \& {Howes}(2013)]{TenBarge:2013a}
{\sc {TenBarge}, J.~M. \& {Howes}, G.~G.} 2013 {Current Sheets and
  Collisionless Damping in Kinetic Plasma Turbulence}. {\em                                                                                                                                                  
  Astrophys.~J.~Lett.\/} {\bf 771}, L27.
  
  \bibitem[{Uritsky} {\em et~al.\/}(2010){Uritsky}, {Pouquet}, {Rosenberg},
  {Mininni} \& {Donovan}]{Uritsky:2010}
{\sc {Uritsky}, V.~M., {Pouquet}, A., {Rosenberg}, D., {Mininni}, P.~D. \&
  {Donovan}, E.~F.} 2010 {Structures in magnetohydrodynamic turbulence:
  Detection and scaling}. {\em Phys.~Rev.~E\/} {\bf 82}~(5), 056326.

\bibitem[{Verniero} {\em et~al.\/}(2018){Verniero}, {Howes} \&
  {Klein}]{Verniero:2017a}
{\sc {Verniero}, J.~L., {Howes}, G.~G. \& {Klein}, K.~G.} 2018 {Nonlinear
  energy transfer and current sheet development in localized Alfv\'en
  wavepacket collisions in the strong turbulence limit}. {\em J.~Plasma                                                                                                                                      
  Phys.\/} In preparation.

\bibitem[{Wan} {\em et~al.\/}(2012){Wan}, {Matthaeus}, {Karimabadi},
  {Roytershteyn}, {Shay}, {Wu}, {Daughton}, {Loring} \& {Chapman}]{Wan:2012}
{\sc {Wan}, M., {Matthaeus}, W.~H., {Karimabadi}, H., {Roytershteyn}, V.,
  {Shay}, M., {Wu}, P., {Daughton}, W., {Loring}, B. \& {Chapman}, S.~C.} 2012
  {Intermittent Dissipation at Kinetic Scales in Collisionless Plasma
  Turbulence}. {\em Phys.~Rev.~Lett.\/} {\bf 109}~(19), 195001.

\bibitem[{Wang} {\em et~al.\/}(2013){Wang}, {Tu}, {He}, {Marsch} \&
  {Wang}]{Wang:2013}
{\sc {Wang}, X., {Tu}, C., {He}, J., {Marsch}, E. \& {Wang}, L.} 2013 {On
  Intermittent Turbulence Heating of the Solar Wind: Differences between
  Tangential and Rotational Discontinuities}. {\em Astrophys.~J.~Lett.\/} {\bf                                                                                                                               
  772}, L14.

\bibitem[{Wu} {\em et~al.\/}(2013){Wu}, {Perri}, {Osman}, {Wan}, {Matthaeus},
  {Shay}, {Goldstein}, {Karimabadi} \& {Chapman}]{Wu:2013}
{\sc {Wu}, P., {Perri}, S., {Osman}, K., {Wan}, M., {Matthaeus}, W.~H., {Shay},
  M.~A., {Goldstein}, M.~L., {Karimabadi}, H. \& {Chapman}, S.} 2013
  {Intermittent Heating in Solar Wind and Kinetic Simulations}. {\em                                                                                                                                         
  Astrophys.~J.~Lett.\/} {\bf 763}, L30.

\bibitem[{Zhdankin} {\em et~al.\/}(2013){Zhdankin}, {Uzdensky}, {Perez} \&
  {Boldyrev}]{Zhdankin:2013}
{\sc {Zhdankin}, V., {Uzdensky}, D.~A., {Perez}, J.~C. \& {Boldyrev}, S.} 2013
  {Statistical Analysis of Current Sheets in Three-dimensional
  Magnetohydrodynamic Turbulence}. {\em Astrophys.~J.\/} {\bf 771}, 124.

\end{thebibliography}
\end{document}